\documentclass[draft,jgrga]{AGUTeX}

\usepackage{lineno}

\usepackage{color}
\usepackage{setspace}

\usepackage{graphicx}
\def\eqlbl#1{\label{eq:#1}}
\def\eqref#1{(\ref{eq:#1})}

\authorrunninghead{FOLEY, BERCOVICI, ELKINS-TANTON}

\titlerunninghead{INITIATION OF PLATE TECTONICS}

\authoraddr{D. Bercovici,
Department of Geology and Geophysics, Yale University, 210 Whitney Avenue, New Haven, CT 06511, USA.
(david.bercovici@yale.edu)}

\authoraddr{L. T. Elkins-Tanton,
School of Earth and Space Exploration, Arizona State University, Tempe, AZ 85287, USA.
(ltelkins@asu.edu)}

\authoraddr{Corresponding Author: B. J. Foley,
Department of Terrestrial Magnetism, Carnegie Institution for Science, 5241 Broad Branch Road NW,
Washington, DC 20015, USA.
(bfoley@dtm.ciw.edu)}

\begin{document}

%
%

\title{Initiation of Plate Tectonics from Post-Magma Ocean Thermo-Chemical Convection}
%

%
%



\authors{Bradford J. Foley, \altaffilmark{1,3}
David Bercovici, \altaffilmark{1} Linda T. Elkins-Tanton \altaffilmark{2}}

\altaffiltext{1}{Department of Geology and Geophysics,
Yale University, New Haven, Connecticut, USA.}

\altaffiltext{2}{School of Earth and Space Exploration, Arizona State University, Tempe, AZ 85287, USA.}

\altaffiltext{3}{Now at: Department of Terrestrial Magnetism, Carnegie Institution for Science,
Washington, DC, USA.}

%
%


\begin{abstract}
Leading theories for the presence of plate tectonics on Earth typically appeal to the role of present day conditions in promoting rheological weakening of the lithosphere.  However, it is unknown whether the conditions of the early Earth were favorable for plate tectonics, or any form of subduction, and thus how subduction begins is unclear.  Using physical models based on grain-damage, a grainsize-feedback mechanism capable of producing plate-like mantle convection, we demonstrate that subduction was possible on the Hadean Earth (hereafter referred to as proto-subduction or proto-plate tectonics), that proto-subduction differed from modern day plate tectonics, and that it could initiate rapidly.  Scaling laws for convection with grain-damage show that, though either higher mantle temperatures or higher surface temperatures lead to slower plates, proto-subduction, with plate speeds of $\approx 1.75$ cm/yr, can still be maintained in the Hadean, even with a CO$_2$ rich primordial atmosphere.  Furthermore, when the mantle potential temperature is high (e.g. above $\approx 2000$ K), the mode of subduction switches to a ``sluggish subduction" style, where downwellings are drip-like and plate boundaries are diffuse.  Finally, numerical models of post-magma ocean mantle convection demonstrate that proto-plate tectonics likely initiates within $\sim100$ Myrs of magma ocean solidification, consistent with evidence from Hadean zircons.  After the initiation of proto-subduction, non-plate-tectonic ``sluggish subduction" prevails, giving way to modern style plate tectonics as both the mantle interior and climate cool.  Hadean proto-subduction may hasten the onset of modern plate tectonics by drawing excess CO$_2$ out of the atmosphere and cooling the climate.   
\end{abstract}

%
%

%

\begin{article}

%
%

\section{Introduction}
\label{sec:intro}

Plate tectonics is the most fundamental process governing the solid Earth, responsible for the observed patterns of seismicity, volcanism, and major geologic features such as mountain belts. In addition, plate tectonics is also crucial for habitability, and possibly even for the origin of life.  Plate tectonics plays a key role in sustaining habitable conditions by buffering atmospheric CO$_2$ through the negative feedbacks involving orogeny, erosion, weathering, and volcanism \citep[e.g.][]{Walker1981}, and was potentially important in the origin of life by providing the energy source for early chemosynthetic organisms \citep{Martin2008}. Despite its importance, the underlying causes of plate tectonics, and why it is absent on other planets such as Mars and Venus, are still a major research topic. Plate tectonics arises from mantle convection via shear localization and weakening of the strong, high viscosity lithosphere, a process which requires an exotic, non-linear rheology (or effective rheology) \citep[e.g.][]{Berco2000,Tackley2000,berco2003}. Over the past two decades significant progress has been made in developing mechanisms capable of producing plate-tectonic style convection and lithospheric weakening \citep[e.g.][]{weinstein1992,berco1993,berco1995a,Moresi1998,trompert1998,Tackley2000a,Korenaga2007,Landuyt2008}, and in hypothesizing why plate tectonics is maintained on Earth but not on other planets \citep[e.g.][]{Moresi1998,Rege2001,Korenaga2007,Lenardic2008,Landuyt2009a,Foley2012}, with most studies pointing to the role of liquid water or a cool climate on Earth.  However, these studies only consider the present day conditions of Earth, leaving open the question of how Earth evolved to this state.  

In particular, grain-damage, a plate generation mechanism (based on grainsize reduction) that captures both dynamic lithospheric weakening and weak-zone memory (a feature that other mechanisms neglect) \citep[e.g.][]{brs1,br4,br6,br2012,br2013,br2014}, predicts that plate tectonics is favored for planets with cool climates, because low surface temperatures suppress grain-growth in the lithosphere and enhance damage \citep{Landuyt2008,Foley2012}.  In addition, low mantle temperatures are more favorable for plate tectonics because they also help suppress lithospheric grain-growth \citep{Foley2012}.  However, the importance of relatively low mantle and surface temperatures, brought about by the influence of lithospheric temperature on grain-growth, is problematic as the early Earth is expected to have a significantly warmer mantle, and potentially a warmer climate.  Given that plate tectonics may be necessary for developing a temperate climate, and is the most efficient way to cool the mantle interior, it is unclear how plate tectonics could ever begin on Earth.  

Another important issue is the nature of plate tectonics on the early Earth.  It is unclear from the geologic record when plate tectonics began, and whether early Earth subduction was even ``plate-tectonic" (in the modern day sense) at all.  Hadean zircons show geochemical signatures consistent with subduction at 4.3-4.4 Ga \citep[e.g.][]{Valley2002,Watson2005,Hopkins2008,Hopkins2010}, though this interpretation is controversial \citep[e.g.][]{Kemp2010}. Furthermore, it is unknown if the Hadean subduction that is potentially indicated by zircon analyses was widespread across the planet, or a local scale (and possibly intermittent) phenomenon.  It is also unclear if other aspects of modern day plate tectonics, such as localized spreading ridges and strike slip faults, or the full Wilson cycle, were concurrent with this possible early Earth subduction. Other studies find that tectonic and petrological indicators of plate tectonics do not appear across the geologic record until $\sim 3$ Ga \citep[e.g.][]{Condie2008,Shirey2011}, and that mantle mixing at this time was less efficient than would be expected from a simple model of plate tectonic style convection \citep[e.g.][]{Debaille2013}.  One possible interpretation of these seemingly contradictory data is that subduction began in the Hadean and Eoarchean, but differed significantly from modern day plate tectonics; for example early Earth subduction may have lacked fully developed ridges or strike-slip faults due to the absence of pre-existing lithospheric weak zones \citep{br2014}.  Furthermore, rheological effects in the mantle and lithosphere could make early Earth subduction more sluggish, and potentially intermittent \citep{Vanhunen2008,Korenaga2011,Vanhunen2012}, explaining some of the geochemical observations \citep{Korenaga2013}.  As such we will refer to subduction in the Hadean, as inferred by our scaling law analysis and numerical models, as proto-subduction or proto-plate tectonics, to distinguish it from modern style plate tectonics which might not develop until much later. 

Finally, most previous studies on the geodynamic feasibility of plate tectonics, or any form of subduction, on the early Earth have focused on the effect a hotter mantle would have on mid-ocean ridge melting and the buoyancy of oceanic lithosphere. Many authors have argued that a thicker, buoyant crust caused by the greater depth of mid-ocean ridge melting could inhibit subduction \citep[e.g.][]{davies1992b,Vlaar1985,Vanthienen2004}, though these studies neglect the eclogite transition in their models (see also reviews by \cite{Davies_treatise} and \cite{Vanhunen2008_review}). In this paper, we focus instead on whether early Earth conditions, specifically high mantle and surface temperatures, allow the effective rheology to produce the requisite lithospheric weakening and shear localization for the operation of plate tectonics.  We therefore neglect chemical buoyancy variations in the slab resulting from mid-ocean ridge melting in this study (the possible effects of crustal buoyancy on our results are discussed in \S \ref{sec:discussion}).  The efficacy of the grain-damage mechanism in forming weak lithospheric shear zones is a fundamentally different constraint than whether the lithosphere could attain negative buoyancy, and one that is an equally significant requirement for subduction.

\subsection{Purpose of This Study}

We put forth a physical model, based on grain-damage, that addresses the problem of initiating early Earth proto-subduction by demonstrating that proto-subduction could be sustained even for the highest plausible Hadean mantle and surface temperatures, and that proto-subduction could initiate rapidly on the early Earth.  Furthermore, we show that proto-subduction was a distinct  style of convection, different in a few key ways from modern day plate tectonics.  We propose that our model is consistent with, and lends geodynamic support to, the inference of Hadean subduction from analyses of ancient zircons. Finally, we posit that the initiation of proto-subduction on the early Earth aids in the development of a temperate, habitable climate, by removing excess CO$_2$ from the primordial atmosphere, and facilitates the onset of modern day plate tectonics by cooling both the mantle and the climate.      
  
Our model is structured in the following manner: First, we use newly developed scaling laws for convection with grain-damage \citep{Foley2013_scaling}, to demonstrate that the early Earth could support convection with proto-subduction, even in the face of high mantle and surface temperatures, by determining the plate speed and heat flux as functions of mantle and surface temperature.  These same scaling laws also show that the style of subduction at these high temperature conditions was different than modern day plate tectonics.  However, the scaling law results only indicate that the steady-state behavior of convection at early Earth conditions would allow proto-subduction; they do not indicate whether proto-subduction could initiate rapidly, within the Hadean or Eoarchean.  We thus use numerical models to constrain how, and how rapidly after magma ocean solidification, proto-subduction initiates on the Earth.  We develop a scaling law for the lag time, or the time it takes for subduction to initiate once solid-state mantle convection has begun (after magma ocean solidification), based on our numerical results; we apply this scaling law to the Hadean Earth to show that the lag time was short, and thus proto-subduction could initiate during the Hadean. The paper is organized as follows: the theory for grain-damage and the governing equations of our numerical convection model are laid out in \S \ref{sec:theory}; scaling laws for convection with grain-damage are used to constrain the operation of proto-subduction on the early Earth in \S \ref{sec:hadean_scale}; the numerical model setup for determining the lag time is outlined in \S \ref{sec:model}, and results are discussed in \S \ref{sec:results}; a scaling law for the lag time is developed in \S \ref{sec:scaling} and applied to the Hadean Earth in \S \ref{sec:app}; finally we discuss our results in \S \ref{sec:discussion} and conclude in \S \ref{sec:conclusions}.
 
\section{Theory}
\label{sec:theory}

Our theoretical formulation of grain-damage was thoroughly explained in \cite{Foley2013_scaling}, so we will only briefly review it here.  We approximate the full, multi-phase theory of grain-damage with Zener pinning and a composite rheology (i.e. one that involves both diffusion and dislocation creep) \citep{br2012}, by a single phase theory that only involves a diffusion creep rheology.  \cite{br2012} show that deformation and damage to the interface between phases (e.g. olivine and pyroxene) combined with pinning effects leads to a state of small-grain permanent diffusion creep, which is observed in natural peridotitic mylonites  \citep{Warren2006}.  We assume that this permanent diffusion creep ``pinned" state prevails throughout the mantle.  In the pinned state, the grainsize of the primary phase is controlled by the curvature of the interface with the secondary phase, and thus damage to the interface leads directly to damage of the bulk phase. This result allows us to reduce the problem to a single phase because the bulk grainsize of the material is determined by the interface curvature; i.e. the interface curvature and grainsize are approximately equal.  We can therefore use the equation governing the interface curvature (equation 4d of \cite{br2012}) to solve for the grainsize directly. 

Assuming that the pinned state prevails throughout the mantle also allows us to neglect non-Newtonian dislocation creep and use a purely grainsize sensitive diffusion creep rheology in our numerical models.  However, in reality the rheology will be controlled by whichever mechanism allows for the easiest deformation \citep[e.g.][]{Rozel2010}, and when grains are large dislocation creep should dominate.  Neglecting dislocation creep in the lithosphere is justifiable because subduction only occurs when damage is effective, therefore resulting in small grain sizes and making diffusion creep the preferred mechanism. The distinction between dislocation and diffusion creep in the mantle interior could in principle influence our results.  However, the relevant physics governing subduction and plate tectonics (or proto-plate tectonics) is controlled by the effective rheology of the lithosphere, and thus is not sensitive to the creep mechanism that predominates in the mantle interior \citep{Foley2013_scaling}; therefore our conclusions should not be significantly impacted by neglecting dislocation creep in the sub-lithospheric mantle. 

\subsection{Damage Formulation}

The viscosity is sensitive to grainsize and temperature as expected for diffusion creep or grain boundary sliding \citep{Hirth2003}:   

\begin{equation}
\eqlbl{eqv}
\mu = \mu_n \exp \left(\frac{E_v}{RT} \right) \left(\frac{A}{A_0} \right)^{-m}
\end{equation}
where $\mu_n$ is a constant, $E_v$ the diffusion creep activation energy ($E_v = 300$ kJ/mol \citep{Karato1993}), $T$ the temperature, $R$ the universal gas constant, $A$ the fineness, or inverse grainsize \citep{br6, Landuyt2008,Foley2012}, and $A_0$ the reference fineness (see Table \ref{tab_param} for a list of key variables and non-dimensional parameters used in this paper). The constant $m$ is equal to 2 for diffusion through the grain (Nabarro-Herring creep) and equal to 3 for diffusion along the grain boundary (Coble creep) \citep[e.g.][]{Evans1995}.  For numerical purposes we use $m=2$, as $m=3$ produces higher degrees of localization and larger viscosity contrasts that can cause numerical convergence problems.  

In the pinned state, fineness is governed by the following evolution equation:

\begin{equation}
\eqlbl{eqd}
\frac{DA}{Dt} = \frac{f}{\gamma}\Psi - h A^p
\end{equation}
where $t$ is time, $f$ is the damage partitioning fraction, which can vary from zero to one, $\gamma$ is the surface free energy, $\Psi$ the deformational work, $h$ the healing rate, and $p$ a constant.  Deformational work is defined as $\Psi = \nabla \underline{v} : \underline{\underline{\tau}}$,  where $\underline{v}$ is the velocity and $\underline{\underline{\tau}}$ is the stress tensor \citep{br6,Landuyt2008}; see also \cite{austin2007}.  

The first term on the right side of \eqref{eqd} represents the partitioning of a fraction ($f$) of deformational work into surface free energy by reducing grainsize (increasing fineness).  The second term on the right side represents reduction of fineness due to normal grain-growth.  The exponent $p$ is approximately 4 in the pinned state \citep{br2012}.  The healing rate constant, $h$, is a function of temperature with an Arrhenius form: 

\begin{equation}
\eqlbl{eqheal}
h = h_n \exp \left(\frac{-E_h}{RT} \right)     
\end{equation}
where $h_n$ is a constant and $E_h$ is the activation energy for grain-growth.  The value of $E_h$ in the pinned state is not well known because most grain-growth experiments are performed with monomineralic samples where pinning effects are not present; this results in low values of $E_h \approx 200$ kJ/mol and rapid grain-growth \citep[e.g.][]{Karato1989}.  In the pinned state, $E_h$ should be higher because grain-growth can only occur through diffusion of material from one grain to another \citep{br2012}.  The kinetics of this process are still uncertain, but some preliminary results indicate lager values of $E_h \approx 400-500$ kJ/mol \citep[e.g.][]{Faul2006,Hiraga2010}.           

\subsection{Governing Equations}

\label{sec:govern_eqs}

We study the initiation of plate tectonics by incorporating grain-damage into a model of infinite Prandtl number, Boussinesq thermo-chemical convection.  The damage formulation, \eqref{eqd}, is non-dimensionalized using the following scales where primes denote non-dimensional variables: $\underline{x} = \underline{x}^{\prime}d$, where $d$ is the depth of the mantle; $t = t^{\prime}d^2/\kappa$, where $\kappa$ is the thermal diffusivity; $\underline{v} = \underline{v}^{\prime}\kappa/d$; $T = T^{\prime}  \Delta T + T_s$, where $\Delta T$ is the temperature difference across the mantle and $T_s$ is the surface temperature; $A = A^{\prime} A_0$; $\underline{\underline{\tau}} = \underline{\underline{\tau}}^{\prime} \mu_m \kappa/d^2$, where $\mu_m$ is the reference viscosity defined at $T_m = \Delta T + T_s$ in the absence of damage; and $E_v = E_v^{\prime} R \Delta T$ and $E_h = E_h^{\prime} R \Delta T$.  The resulting non-dimensional equation governing fineness evolution is: 

\begin{equation}
\eqlbl{eqdam}
\frac{DA'}{Dt'} = D \psi  \exp{\left(\frac{E_v'}{T'+T_s^*} - \frac{E_v'}{1+T_s^*}\right)} A'^{-m} - H\exp{\left(\frac{-E_h'}{T'+T_s^*} + \frac{E_h'}{1+T_s^*}\right)} A'^p 
\end{equation}
where $\psi = \nabla \underline{v}' : ( \nabla \underline{v}' +  (\nabla \underline{v}')^T)$, $T_s^* = T_s/\Delta T$, $D$ is the non-dimensional damage number and $H$ is the non-dimensional healing number.  These quantities are defined as $D = f\mu_m\kappa/(\gamma A_0 d^2)$ and $H = h_m A_0^{(p-1)}d^2/\kappa$ where $h_m = h(T_m)$.  

The equations for conservation of mass, momentum, energy, and composition, expressed in terms of non-dimensional variables using the same scales as above are:  

\begin{equation}
\eqlbl{eqmass}
\nabla \cdot \underline{v}' = 0
\end{equation}
\begin{equation}
\eqlbl{eqmom}
0=-\nabla P' + \nabla \cdot (2 \mu' \dot{\underline{\underline{\varepsilon}}'}) + Ra_0(T' - BC') \hat{\underline{z}}
\end{equation}
\begin{equation}
\eqlbl{eqenergy}
\frac{\partial{T'}}{\partial{t'}}+ \underline{v}' \cdot \nabla T' = \nabla^2 T' +  Q'
\end{equation}
\begin{equation}
\eqlbl{eqchem}
\frac{\partial{C'}}{\partial{t'}}+ \underline{v}' \cdot \nabla C' = \frac{1}{Le} \nabla^2 C'
\end{equation}
where $P'$ is the non-hydrostatic pressure, $\dot{\varepsilon_{ij}}' = (\partial{v'_i} / \partial{x'_j} + \partial{v'_j} / \partial{x'_i} )/2$ is the strain rate, $\hat{\underline{z}}$ is the unit vector in the vertical direction, and $Ra_0$ is the reference Rayleigh number; $Ra_0 = (\rho \alpha g \Delta T d^3)/(\kappa \mu_m)$ where $\rho$ is density, $\alpha$ is thermal expansivity, and $g$ is acceleration due to gravity. The non-dimensional internal heating rate, $Q'$, is defined as $Q' = Q d^2/(\Delta T \kappa)$, where $Q = Q_v/(\rho C_p)$, $Q_v$ is the volumetric heating rate (with units of Wm$^{-3}$), and $C_p$ is the specific heat. The composition, $C'$, represents heavy element concentration, such that $C' = 1$ is chemically dense, and the buoyancy is set by the buoyancy number, $B = \beta \Delta C / (\alpha \Delta T)$, where $\beta$ is the chemical expansivity.  The buoyancy number thus gives the ratio of the compositional density difference to the (non-adiabatic) thermal density difference across the mantle.  The Lewis number, $Le= \kappa/\kappa_C$, is the ratio of the thermal diffusivity to the chemical diffusivity. Finally, we define parameters to describe the variation of viscosity and healing across the mantle due to temperature dependence; $\mu_l' = \mu_l/\mu_m$, the viscosity ratio in the absence of damage ( i.e. with $A = A_{ref}$), and $h_l' = h_l/h_m$, the healing ratio, where the subscript $l$ denotes the value in the lithosphere (i.e. at $T'=0$ for all numerical models).  The viscosity and healing ratios are therefore equivalent to the non-dimensional viscosity and healing, respectively, at $T' = 0$.    

\begin{table}
\caption{Table of Variables and Non-Dimensional Parameters}
\label{tab_param}
\begin{tabular}{c c c}
\hline
Variable & Definition & Reference Equation \\ \hline
$A$ & Fineness (inverse grain-size) & \eqref{eqd} \\
$\mu$ & Viscosity & \eqref{eqv} \\
$h$ & Healing or grain-growth rate & \eqref{eqheal} \\
$E_v$ ($E_h$) & Activation energy for viscosity (healing) & \eqref{eqv} (\eqref{eqheal}) \\ 
$f$ & Damage partitioning fraction & \eqref{eqd} \\ 
$\underline{v}$ & Velocity & \eqref{eqmass}, \eqref{eqmom} \\  
$P$ & Non-hydrostatic pressure & \eqref{eqmom} \\  
$T$ & Temperature & \eqref{eqmom}, \eqref{eqenergy} \\  
$Q$ & Internal heating rate & \eqref{eqenergy} \\ 
$C$ & Composition & \eqref{eqmom}, \eqref{eqchem} \\  
\hline
Non-Dimensional Parameter & Definition & Reference Equation \\ 
\hline
$Ra_0$ & Reference Rayleigh number & \eqref{eqmom} \\
$B$ & Buoyancy number & \eqref{eqmom} \\
$Le$ & Lewis number & \eqref{eqchem} \\
$D$ & Damage number & \eqref{eqdam} \\
$H$ & Healing number & \eqref{eqdam} \\
$\mu_l/\mu_m$ ($h_l/h_m$) & Viscosity (healing) ratio across the mantle & - \\
$m$ & Grainsize sensitivity exponent for viscosity $(m=2)$ & \eqref{eqv} \\
$p$ & Grain-growth exponent $(p=4)$ & \eqref{eqd} 
\end{tabular}      
\end{table}

\section{Hadean Proto-subduction}
\label{sec:hadean_scale}

We use scaling laws describing the plate velocity and heat flow for statistically steady-state convection with grain damage to estimate how these quantities scale with increasing mantle and surface temperature, and thus assess whether lithospheric recycling could be maintained at the high mantle temperatures, and possibly high surface temperatures, of the early Earth. Our results in this section give the expected plate speed and heat flow for mantle convection at a given mantle and surface temperature, but do not treat thermal evolution; however, the implications of our results for the thermal evolution of the mantle are discussed in \S \ref{sec:scale_results}. 

\subsection{Scaling Laws for Convection with Grain-Damage}
\label{sec:scale_model}

As shown by \cite{Foley2013_scaling}, convection with grain-damage displays both mobile lid and stagnant lid behavior. The mobile lid behavior is governed by two separate sets of scaling laws that describe two distinct regimes of convection: the transitional regime, where plate speed is dictated by the viscous resistance of lithospheric shear zones; and the fully mobile regime, where lithospheric shear zones have been so extensively weakened by damage that they no longer significantly resist subduction. In the fully mobile regime the surface is so thoroughly broken up by damage that convection is no longer plate-like \citep{Foley2013_scaling}; thus, plate-tectonic style convection occurs in the transitional regime. As a result, we can assume that Earth is in the transitional regime, at least for the present day mantle interior and surface temperatures, and neglect the fully mobile regime when applying our scaling laws to the Earth. In addition, a non-plate-tectonic style of mobile lid convection, referred to as ``sluggish subduction," also occurs within the transitional regime, near the boundary with the stagnant lid regime. The sluggish subduction style is characterized by drip-like, rather than slab-like, downwellings and diffuse plate boundaries, a result of less effective damage leading to higher viscosity lithospheric shear zones, which resist plate motion. However, there are still appreciable surface velocities, and rates of lithospheric recycling into the mantle, when subduction is sluggish \citep{Foley2013_scaling}.

Scaling laws for mantle convection with grain-damage were developed in \citep{Foley2013_scaling} for bottom heated convection, while the Earth is predominantly internally heated. However, applying the bottom heated scaling laws to the Earth only introduces a small error, because convecting systems with different heating modes are generally found to follow scaling laws with the same form when the system is analyzed in terms of the internal Rayleigh number, $Ra_i$, i.e. the Rayleigh number where the temperature drop and viscosity are defined at the average interior temperature of the mantle \citep{Reese1999,Solomatov2000b,Korenaga2009,Korenaga2010}. Differences in heating mode have only been found to change the scaling law exponents by up to $\approx 10$ \% \citep{Moore2008}; changes of this magnitude to our scaling law exponents would alter the exact values of plate speed and heat flow we calculate for the early Earth, but would not impact our overall results. Strictly speaking, the scaling laws from \cite{Foley2013_scaling} are developed in terms of the reference Rayleigh number, $Ra_0$; however, the numerical models used to develop these scaling laws have asymmetrical boundary layers, where the majority of the temperature drop across the mantle ($\Delta T$) is accommodated by the temperature difference over the top thermal boundary layer, leaving a much smaller temperature difference over the weaker bottom boundary layer.  This means that the interior mantle temperature, $T_i$, is approximately equal to the basal mantle temperature, $T_m$, and the reference Rayleigh number, $Ra_0$, is approximately equal to the internal Rayleigh number, $Ra_i$.  Thus our scaling laws developed from basally heated numerical models can be applied to the internally heated early Earth (further justification for this assumption can be seen in \S \ref{sec:results_times}, where internally and basally heated models produce nearly identical timescales for initiating mobile lid convection).       

With the assumption that $T_i \approx T_m$ and $Ra_i \approx Ra_0$ in the numerical models of \cite{Foley2013_scaling}, the transitional regime scaling law for heat flow is, 

\begin{equation}
\eqlbl{Nu_trans} 
q = a \left( \frac{k (T_i-T_s) }{d} \right) \left(\frac{L}{d}\right)^{-\frac{1}{10}} \left(\frac{\mu_l}{\mu_i}\right)^{-\frac{1}{4}} \left(\frac{Dh_m \mu_i}{Hh_l \mu_m} \right)^{\frac{1}{3}} Ra_i^{\frac{2}{3}}
\end{equation}
where $k$ is thermal conductivity, $L$ is the aspect ratio of convection cells (or the plate length), and $\mu_l$ and $h_l$ are defined at the lithosphere temperature, $T_l$, which is defined below (\S \ref{sec:Tl_param}).  As stated above, $T_i$ represents the average potential temperature of the well mixed mantle interior (i.e. the isothermal core of a convection cell), $\mu_i$ is the undamaged viscosity at $T=T_i$, and $Ra_i$ is the internal Rayleigh number.  The factor of $\mu_i/\mu_m$ that appears in the damage to healing ratio term scales the damage number to the interior mantle viscosity, and the factor $h_m/h_l$ scales the healing number to the lithospheric healing rate. The plate velocity in the transitional regime scales as         

\begin{equation}
\eqlbl{vl_trans}
v_l = b \left(\frac{\kappa}{d}\right) \left(\frac{L}{d}\right)^{\frac{4}{5}} \left(\frac{\mu_l}{\mu_i}\right)^{-\frac{1}{2}} \left(\frac{Dh_m \mu_i}{Hh_l \mu_m} \right)^{\frac{2}{3}} Ra_i^{\frac{4}{3}} 
\end{equation}
where $a$ and $b$ are constants ($a\approx 0.04$ and $b\approx 0.0025$).

We define the boundary between plate-tectonic style convection and the non-plate-tectonic sluggish subduction style of convection using the criterion from \cite{Foley2013_scaling}, that the plate velocity must be at least 10 \% of the interior mantle velocity for plate-like convection. \cite{Foley2013_scaling} defined the interior mantle velocity, $v_m$, as the horizontal velocity at the base of the mantle; however, the temperature drop across the bottom boundary layer for the early Earth is not well known, so we instead define $v_m$ as the velocity of a drip off the base of the lithosphere. Sub-lithospheric drips are driven by a smaller temperature difference than that across the lithosphere ($T_i - T_s$). The temperature difference driving sub-lithospheric drips can be approximated by the rheological temperature difference for stagnant lid convection, $\Delta T_{rh}$, which defines the temperature contrast for the lower, mobile portion of the lid.  The interior mantle velocity is then given by the scaling law for velocity in the stagnant lid regime \citep{Slava1995}, 

\begin{equation}
\eqlbl{vm}
v_m = \frac{c\kappa}{d} \left(\frac{Ra_i}{\theta}\right)^{\frac{2}{3}} 
\end{equation}      
where $c$ is a constant of order 0.1, and $\theta$ is the Frank-Kamenetskii parameter, defined as 

\begin{equation}
\theta = \frac{E_v (T_i - T_s)}{RT_i ^2} 
\end{equation}
\citep{Korenaga2009}.  We approximate the sub-lithospheric mantle as a simple, Newtonian fluid with a temperature dependent viscosity, because the high mantle temperatures in the Archean and Hadean would likely suppress damage in the mantle via rapid grain-growth.  Strictly speaking, rapid grain-growth activates the non-Newtonian dislocation creep mechanism.  However, dislocation creep can be well approximated by a Newtonian fluid with the proper choice of activation energy, that is fortuitously approximately equal to the diffusion creep activation energy of $\approx 300$ kJ/mol \citep{christensen1984c}. 

\subsection{Lithosphere Model and Assumed Parameters}
\label{sec:Tl_param}

The final definition needed to apply our grain-damage scaling laws to Hadean mantle convection is the lithospheric temperature, $T_l$. The definition of $T_l$ is based on the lithospheric strength profile, and was fully explained in \cite{Foley2012}; for this reason we will only give a brief overview here. Our formulation conservatively assumes that viscous flow and damage must account for deformation all the way to the base of the frictional sliding layer (i.e. where frictional sliding gives way to semi-brittle, semi-ductile deformation), thereby including the region of peak-strength in the mid-lithosphere \citep[e.g.][]{Kohlstedt1995}. Lithospheric temperature is therefore defined as the temperature at the transition from frictional sliding to semi-ductile deformation. As shown by \cite{Foley2012}, a simple, conservative approximation for the transition from frictional sliding to semi-ductile deformation is to assume that it occurs at a fixed confining pressure; including the influence of temperature on this transition point expands the frictional sliding region and therefore positively biases the effectiveness of grain-damage. The base of the frictional sliding layer is constrained by the depth of faulting in oceanic lithosphere, which is approximately 20 km on Earth \citep[e.g.][]{Bergman1988,Tichelaar1993}, corresponding to a confining pressure of $P_f \approx 600$ MPa. Assuming a linear temperature profile in the lithosphere, we write

\begin{equation}
\eqlbl{lith_temp}
T_l = \left(\frac{T_i - T_s}{\delta_l}\right) \left(\frac{P_f}{\rho_0 g}\right) + T_s 
\end{equation}      
where $\rho_0$ is the lithosphere density ($\approx 3000$ kg/m$^3$).  The lithosphere thickness, $\delta_l$, is determined from the heat flow scaling law \eqref{Nu_trans} using $\delta_l = (T_i - T_s)/q$. In the transitional regime $\delta_l$ is itself a function of $T_l$, and \eqref{lith_temp} must be solved numerically.   

Though simple, our model for the lithosphere reproduces the important aspects of the lithospheric strength profile.  If the lithosphere is thinner (i.e. $\delta_l$ is smaller), the lithosphere temperature increases because the transition from frictional sliding to semi-brittle/semi-ductile deformation now occurs at a higher temperature. Therefore the viscosity ratio decreases (owing to a smaller lithospheric viscosity) and the effective strength of the lithosphere, before grainsize effects are included, decreases as well. The same effect occurs for a hotter mantle; the temperature difference across the lithosphere, $T_i - T_s$, increases, thereby increasing $T_l$ and dropping the effective lithospheric strength, again before any effects of grainsize are included.

The healing term is constrained by comparing the theoretical formulation of grain-damage with Zener pinning to grain-growth experiments \citep{br2012,br2013}.  As in \cite{Foley2013_scaling}, we assume an activation energy for grain-growth of $E_h = 500$ kJ/mol, which is consistent with grain-growth experiments that consider the effects of secondary phases \citep[e.g.][]{Evans2001,Faul2006,Hiraga2010}.  We also calculate the constant $h_n$ in the same manner as \cite{Foley2013_scaling} (see Section 8.2), and find $h_n = 2 \times 10^{-7}$ m$^3$/s for $p=4$.  We use a value of $f \approx 10^{-9}$ to match Earth's current day plate speed and heat flow with $L/d = 1.5$, the average value from the numerical models in \cite{Foley2013_scaling}.  We also assume the mean mantle density $\rho=4000$ kg/m$^3$, $g = 10$ m/s$^2$, $\alpha = 3 \times 10^{-5}$ 1/K, $d=2890$ km, $\kappa = 10^{-6}$ m$^2$/s, $E_v = 300$ kJ/mol, $A_0 = 1000$ 1/m (1 mm reference grainsize), and $\mu_n = 2\times 10^{11}$ Pa s, which results in a reference mantle viscosity of $10^{21}$ Pa s.  

\subsection{Results}
\label{sec:scale_results}

\begin{figure}
\includegraphics[scale = 0.6]{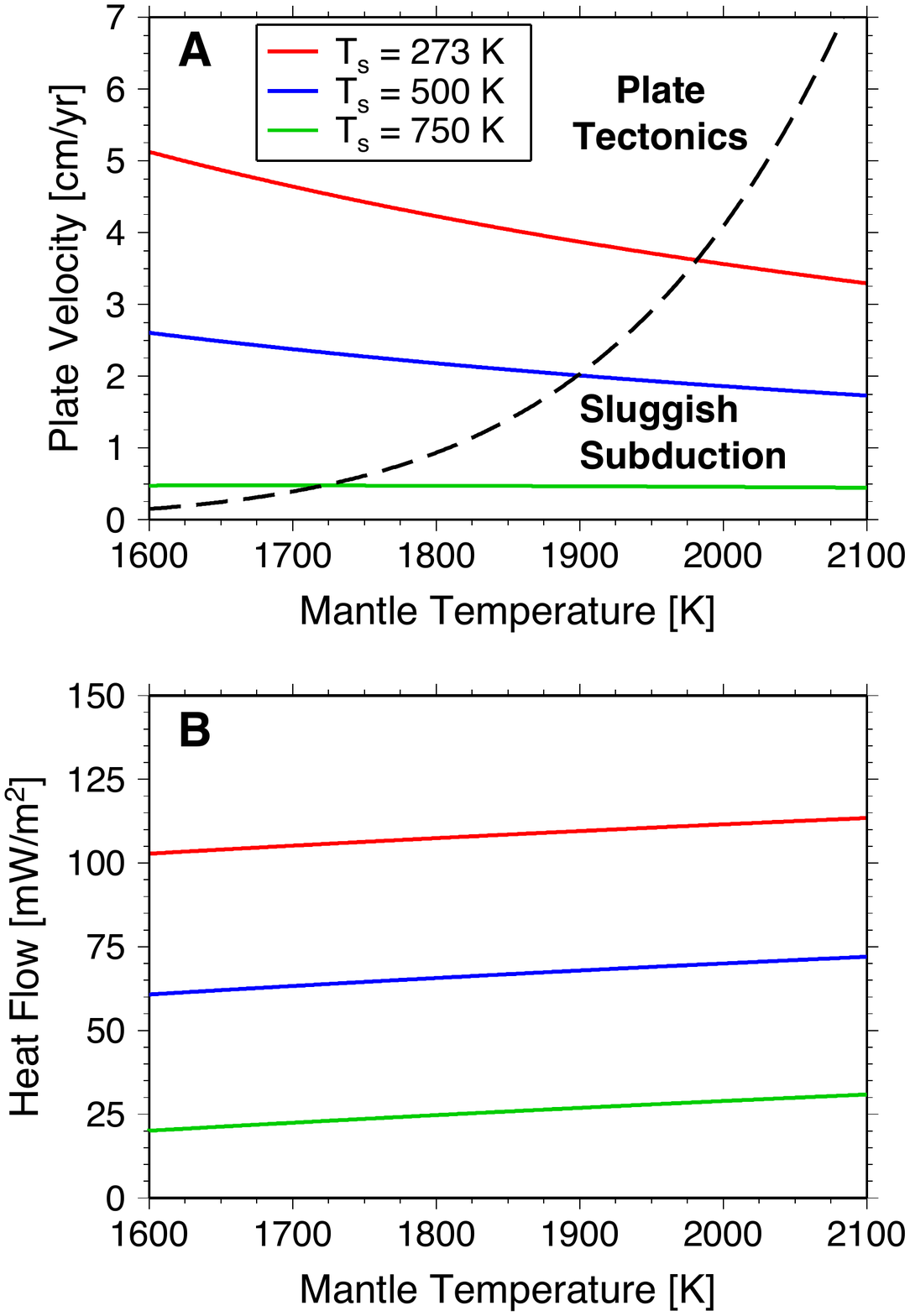}
\caption{\label {fig:hadean_scale} Plots of plate velocity (a) and heat flow (b) versus mantle potential temperature, at three different surface temperatures: $T_s = 273$ K (red line), $T_s = 500$ K (blue line), $T_s = 750$ K (green line). The dashed black line shows $v_m/10$ (Equation \eqref{vm}), giving the boundary between the nominally plate-tectonic regime and a non-plate-tectonic, ``sluggish subduction" regime.}  
\end{figure}

Plate speed decreases with increasing mantle potential temperature (Figure \ref{fig:hadean_scale}A) as a result of a competition between the influence of warmer mantle temperatures on the effective viscosity of lithospheric shear zones, and on the Rayleigh number.  Warmer mantle temperatures tend to produce higher lithosphere temperatures by increasing $T_i - T_s$ (see \eqref{lith_temp}), thereby reducing the viscosity of the undamaged lithosphere. However, the warmer lithosphere temperature, $T_l$, also increases the lithospheric healing rate substantially (significantly lowering the damage to healing ratio), which acts to increase viscosity.  The effect of more rapid grain-growth dominates over the direct temperature effect on lithospheric viscosity, mainly because $E_h$ is larger in magnitude than $E_v$, and lithospheric shear zones thus become effectively stronger at warmer mantle temperatures.  The increase in the effective viscosity of lithospheric shear zones then competes with the effect warmer mantle temperatures have on the internal Rayleigh number, $Ra_i$, as warmer mantle temperatures increase $Ra_i$ by dropping the interior mantle viscosity and increasing the temperature difference driving convection.  However, the strengthening of lithospheric shear zones brought about by higher healing rates wins out, and the end result is the decrease in plate speed shown in Figure \ref{fig:hadean_scale}A. We note that this increase in viscosity due to faster grain-growth only occurs up to a critical grainsize, above which dislocation creep ensues; however, this transition does not take place until convection is in the stagnant lid regime, and is thus not important to our results. 

The decrease in plate speed with increasing mantle temperature is small; at Earth's present day surface temperature of 273 K, plate speed only decreases from $\approx 5.25$ cm/yr at $T_i = 1600$ K to $\approx 3.25$ cm/yr at $T_i = 2100$ K, which is near the point where the whole mantle would melt (at these high mantle temperatures, significant upper mantle melting will occur; the possible effects of this are discussed in \S \ref{sec:discussion}).  If the climate is hotter, which is possible in the Hadean, then the plate speed is lower still.  At a surface temperature of 500 K, possible for a warm greenhouse climate \citep{Zahnle2007}, plate speed is $\approx 1.75$ cm/yr at $T_i = 2100$ K, while at a Venusian surface temperature of 750 K, plate speed is very low ( $\approx 0.4$ cm/yr).  For all three surface temperatures, the plate speed drops below the threshold for nominally plate-tectonic style convection at high mantle temperatures.  High mantle temperatures increase the interior mantle velocity significantly; combined with the fact that plate speed decreases when interior mantle temperature increases, the plate speed becomes too small a fraction of the mantle velocity for convection to be considered plate-like.  Thus at high mantle temperatures (at or above $\approx 2000$ K for a surface temperature of 273 K) subduction would be drip-like, and plate boundaries diffuse.  However, there is still appreciable lithospheric recycling into the mantle, except for the case with a Venusian surface temperature where plate speed is very low.      

The heat flux increases with mantle temperature, but only very moderately (Figure \ref{fig:hadean_scale}B).  This increase is due to the competing effects mantle temperature has on the thickness of the lithosphere, $\delta_l$, and on the temperature drop across the lithosphere.  The lithospheric thickness increases with $T_i$ for the same reason that $v_l$ decreases; higher lithospheric healing due to the warmer mantle.  A thicker lithosphere acts to decrease the heat flux.  However, the temperature drop across the lithosphere also increases with larger $T_i$, acting to increase the heat flux.  The end result, when these two effects are combined, is the observed weak increase in heat flux.  Higher surface temperatures, however, cause lower heat flows.  The primary cause is that the lithosphere is thicker owing to more rapid grain-growth; reinforcing this effect is that the temperature drop across the lithosphere, $T_i - T_s$, also decreases when surface temperature is higher.    

Our results show that increasing mantle temperature alone can not shut down proto-subduction by limiting the effectiveness of lithospheric shear localization with grain-damage; as long as mantle temperatures are cool enough for the magma ocean to have solidified, proto-subduction is geodynamically favorable (given a cool surface temperature).  Our results also indicate a relatively low heat flow for the Hadean due to the sluggish proto-subduction caused by rapid lithospheric healing. This low heat flow could have important implications for the thermal evolution of the early Earth and its primary heat loss mechanism.  In particular, an important question is whether the convective heat flow we estimate can remove the radiogenic heating present just after magma ocean solidification.  At $T_s = 273$ K and $T_i = 2100$ K we estimate a heat flow of $\approx 115$ mWm$^{-2}$, corresponding to $\approx 60$ TW of total heat loss, assuming no continental crust.  If the surface temperature is 500 K, heat flow is only $\approx 75$ mWm$^{-2}$ and the total convective heat loss is $\approx 38$ TW.  The radiogenic heat production in the Hadean depends on the current day radiogenic heating, which is estimated to be $\approx 15 - 20$ TW, lumping together both mantle and crustal heat production because the Hadean Earth may have lacked continental crust \citep[e.g.][]{korenaga2006,Jaupart_treatise}.  Projecting these estimates back to the Hadean using the procedure from \cite{korenaga2006} gives $\approx 70-100$ TW, which is greater than our convective heat loss.  However, if some heat producing elements are sequestered in a primordial crust, the mantle heat production would be lower.  Furthermore, significant upper mantle melting and subsequent volcanism would transport a large amount of heat out of the mantle (as much as 2-3 times the convective heat flux) and could likely make up for this difference as well \citep{Nakagawa2012}. Nevertheless, future studies are needed to more fully understand heat transport by melting, and the effect this would have on plate-generation with grain-damage and the thermal evolution of the early Earth, as these topics are beyond the scope of this study.

Our results are robust to uncertainties in the parameters for damage and healing.  The results are not strongly sensitive to $f$; for example a factor of 100 increase in $f$ only increases the plate speed at present day conditions by 2 cm/yr.  Our value of $f \approx 10^{-9}$ is lower than some estimates based on geological observations and mineral physics experiments, which indicate $f \sim 10^{-4}$ at a temperature of approximately 1000 K \citep{austin2007,Rozel2010}. However, even if $f$ were higher, our results that Hadean mantle convection could support at least some form of lithospheric recycling would still hold (in fact proto-subduction would be even more likely with a larger $f$). The activation energy for healing, $E_h$, is also poorly known \citep[e.g.][]{Evans2001}.  However, we find that our results are largely insensitive to plausible variations of $E_h$ from $\approx 300-600$ kJ/mol, because of a negative feedback between lithosphere temperature and thickness.  From \eqref{Nu_trans}, larger $E_h$ would make lithospheric thickness increase more rapidly with increasing lithosphere temperature. However, as shown by \eqref{lith_temp}, $T_l$ depends inversely on the thickness of the lithosphere.  The end result when these two competing effects are combined, is that varying $E_h$ has only a minor influence on how $\delta_l$, and thus heat flow and plate speed, scales with mantle (or surface) temperature. 

\subsection{Summary of Scaling Law Results}
\label{sec:scale_disc}

Our results indicate that proto-subduction could be maintained in the Hadean as long as Earth's climate was cooler than present day Venus.  The climate of the early Earth is not well constrained; however combining Hadean zircon data with simple early Earth climate models we can estimate the maximum plausible surface temperature the early Earth could have faced.  A model of the post magma ocean Earth's climate finds a surface temperature of $\approx 500$ K if most of Earth's present day inventory of CO$_2$ was in the atmosphere \citep{Zahnle2007}.  This is consistent with observations from ancient zircons, which imply the surface temperature could not have exceeded $\approx 200^{\circ}$ C in the Hadean \citep{Valley2002}.  Thus, surface temperatures were not high enough to eliminate proto-subduction, with plate speeds of at least 1.75 cm/yr, on the early Earth. However, the style of subduction at early Earth conditions was likely not plate-tectonic due to the effects of more rapid lithospheric grain-growth.  Hadean proto-subduction would be more drip-like, rather than slab-like, and possibly more intermittent as well, while plate boundaries would be broad and diffuse.  Though this style of convection is not strictly plate-tectonic, it still allows for significant recycling of surface materials into the mantle and vice-versa, which is the primary role plate tectonics plays in regulating climate.  However, the scaling laws used to predict Hadean proto-subduction are based on convection in steady-state.  This analysis does not predict how proto-plate tectonics initiates, or whether initiation can occur rapidly enough for proto-subduction to occur by the end of the Hadean (i.e., the time for convection to reach a proto-plate-like steady-state is unknown). To address the initiation time for proto-subduction we defer to numerical analysis, which is discussed next.

\section{Numerical Model Setup}
\label{sec:model}

We use numerical convection models whose initial conditions are representative of the mantle immediately following magma ocean solidification to assess how quickly proto-subduction (or proto-plate tectonics) can initiate on the early Earth. In particular, we are interested in the evolution from the beginning of solid state mantle convection to the onset of proto-subduction, (or mobile lid convection, as it can be more generally described when discussing the numerical models) and how the timescale for initiating mobile lid convection depends on damage, healing, Rayleigh number, and viscosity ratio.  We measure the timescale for initiating mobile lid convection in our models as the lag time between when convection begins and when subduction first initiates; this ensures that the timing of the first convective overturn, which depends heavily on the initial condition, does not influence the results. The hypothesized moon forming impact likely melted a large portion of the planet \citep{Canup2004}, and the resulting silicate vapor atmosphere was hot enough to melt any areas of the surface that survived the impact, resulting in a global magma ocean \citep[e.g.][]{Pahlevan2007,Zahnle2007}.  We make the simplifying assumption that the mantle never experienced total or near total melting after the post-impact magma ocean solidified, and thus treat the newly solidified mantle, after the moon forming impact, as time zero for mantle convection and the initiation of proto-subduction.

Our goal with the numerical models is not to replicate the post-impact Earth; the high Rayleigh number and very high viscosity ratio of the newly solidified mantle is not numerically tractable.  Instead, we seek to understand how the parameters governing convection with grain-damage, e.g. the Rayleigh number, viscosity ratio, damage to healing ratio, internal heating rate, etc., influence the timescale for initiating mobile lid convection by systematically varying these quantitates in our numerical models.  We then use the numerical results to construct a simple scaling law that allows us to extrapolate our model results to Hadean Earth conditions, and estimate the time required to initiate proto-subduction.  As such, the initial conditions for our models, outlined in \S \ref{sec:init_conditions}, are only intended to be a general representation of the post-magma ocean mantle, and are not specific in terms of quantities such as mantle temperature; specific mantle temperatures are chosen when applying our scaling law to the Hadean. 

Some recent studies with the pseudoplastic rheology have found a hysteresis, where models with the same parameters can display different tectonic regimes (e.g. stagnant lid versus plate tectonics) depending on whether the initial condition was stagnant lid or mobile lid \citep{Crowley2012,Weller2012}.  In particular, they find that starting from a stagnant lid initial condition impedes the development of mobile lid convection.  We have not observed a similar effect with grain-damage, as illustrated by the fact that our models develop mobile lid convection within the same parameter space as other studies that utilize different initial conditions \citep{Foley2012,Foley2013_scaling}.  The reason why our models with grain-damage do not to display the same hysteresis effects as those in \cite{Weller2012} likely comes down to differences in the plate generation mechanism, i.e. grain-damage versus pseudoplasticity (see \S \ref{sec:comp_previous} for a more detailed discussion).   Furthermore, the linear temperature profile we use as an initial condition results in a high viscosity lid.  As a result, our numerical models typically evolve through a period of stagnant lid convection before subduction initiates; thus any hysteresis caused by our choice of initial condition would make the development of proto-plate tectonics more difficult, and does not bias our result.

As we vary the parameters governing damage and convection over a wide range, not all of the numerical results will fall in the sluggish subduction regime as described in \S \ref{sec:scale_model}.  However, the lag time for the initiation of proto-plate tectonics scales consistently regardless of whether the ensuing steady-state behavior is in the nominally plate-tectonic or sluggish subduction regimes, so using all of our numerical results to constrain the scaling law for the lag time is justifiable. It comes as no surprise that the lag time scales consistently for either style of subduction considering both styles follow the same scaling laws for plate speed and heat flow. 

\subsection{Initial Conditions}    
\label{sec:init_conditions}

We constrain the initial conditions of our convection calculations from magma ocean solidification models.  An important condition is the initial temperature profile.  If magma ocean solidification occurs more rapidly than the initial overturn time for thermal convection in the growing solid layer, then the solidified mantle will follow the solidus temperature \citep{Lindy2003,Lindy2008}.  However, if solid-state convection begins before solidification is complete, then some portion of the lower mantle, which solidifies first, will follow an adiabat \citep{solomatov2000}.  As either scenario leads to a large temperature difference across at least some portion of the mantle, we assume a solidus temperature profile throughout the mantle as this scenario can be approximated in a simple manner in the numerical models.    

The top and bottom boundary conditions are isothermal, with $T=T_m$ at the base of the mantle and $T=T_s$ at the surface.  To approximate the solidus temperature profile in the mantle interior, we assume that temperature decreases linearly from $T_m$ at the base of the mantle, to $T_{sol0}$, the solidus temperature at zero pressure, just beneath the surface. Thus, at time zero, the temperature at the surface is a step function from $T_s$ to $T_{sol0}$.  We use a value of 0.2 for the non-dimensional zero pressure solidus temperature for all models, based on the solidus temperature profile from \cite{Herzberg1996}. For the models including internal heating, the bottom boundary is switched to a no flux condition, so that these models are purely internally heated (the initial temperature profile is the same as the purely bottom heated cases).  A test case with both bottom and internal heating (at $Q' = 100$) finds that the lag and onset times only change by $\approx 3$ \% as compared to the corresponding purely internally heated model.
 
We next consider whether chemical differentiation occurs during magma ocean solidification.  Chemical differentiation occurs if crystals become segregated from the liquid, by settling at the base of the magma ocean and expelling interstitial melt during compaction, and are unable to maintain chemical equilibrium \citep[e.g.][]{Lindy2003,solomatov2000}.  \cite{solomatov2000} estimates that the typical crystal size in the magma ocean is approximately equal to the critical size needed to keep crystals suspended in the convecting liquid, and thus maintain chemical equilibrium between the crystals and the melt. Given the uncertainties inherent in the estimate of \cite{solomatov2000}, whether chemical differentiation occurs during solidification is unknown. If crystals do settle out of the melt and fractionation occurs, the melt becomes enriched in heavy elements (e.g. iron), causing late-crystallizing minerals to be compositionally dense. As magma ocean crystallization generally precedes from bottom up, late crystallizing minerals form near the surface of the now solid mantle, producing a gravitationally unstable compositional density profile that will overturn \citep{Hess1995,solomatov2000,Lindy2003,Lindy2008}. Overturn is unlikely to occur until after the entire magma ocean has solidified because the overturn timescale depends strongly on the depth of the solid layer, and only becomes comparable to the solidification timescale when the solid layer is approximately as thick as the mantle \citep{Lindy2003}.  

Given the uncertainty in chemical differentiation, we run models both with and without chemical heterogeneity.  For the models including chemical heterogeneity, we assume compositional density decreases linearly with depth, and that there is zero compositional flux at the top and bottom boundaries.  Based on the results of \cite{Lindy2008}, the Buoyancy number, $B$, is of order 1 and we use this value as a base line; however, we run some cases varying $B$ from 0.5-3 to constrain the influence of this parameter.  The Lewis number is expected to be large because chemical diffusion is much slower than thermal diffusion; we use $Le = 10^4$ as a baseline, and run some cases at higher $Le$ to confirm that $Le = 10^4$ does not  significantly affect our results.         

We assume an initially static mantle, which is expected for either model of magma ocean solidification (even in \cite{solomatov2000}, solid-state convection ceases when the solidification front reaches the surface).  We also assume a uniform initial fineness of $A=A_0$, i.e. an initially undamaged mantle, because the mantle has just solidified (see Figure \ref{fig:init_pform}A for the initial condition used for all thermo-chemical models). We use a small, random number perturbation to the initial compositional field (or thermal field when composition is not included) in the lower mantle at $z' = 0.25$; the amplitude of this perturbation is set to $5 \times 10^{-3}$ for most models, and varied from $10^{-5} - 0.2$ in one suite of models designed to test the influence of the initial perturbation. A random number sequence excites instability at all wavelengths larger than the grid resolution and less than the horizontal aspect ratio of the model domain, which is $4 \times 1$.  Both of these limits are far from the most unstable mode for Rayleigh-Taylor instabilities \citep[e.g.][]{turc1982}, so the frequency content of our initial perturbation does not impact the results in any significant way.    

We solve the coupled damage and convection equations \eqref{eqdam}-\eqref{eqchem} using the finite-volume code described in \cite{Foley2013_scaling}.  The only difference is the additional advection-diffusion equation for composition, which we solve using the same methods used for temperature, as outlined in \cite{Foley2013_scaling}: we use the non-oscillatory version of MPDATA for advection \citep{Smolark1984,Smolark1990} and a Crank-Nicholson time discretization for the diffusion term.  All experiments for this study were performed in a $4 \times 1$ domain, with a typical resolution of $512 \times 128$; however a finer grid of $1024 \times 256$ is used for models with high $Ra_0$ or $D$. We perform a large suite of models varying $D$, $Ra_0$, $\mu_l/\mu_m$, $h_l/h_m$, $Q'$, $Le$, and $B$ to determine how the initiation time for plate tectonics scales with these key parameters.           

\section{Numerical Results}
\label{sec:results}

\subsection{General Evolution of Convective Planform}

\begin{figure}
\includegraphics[scale = 0.7]{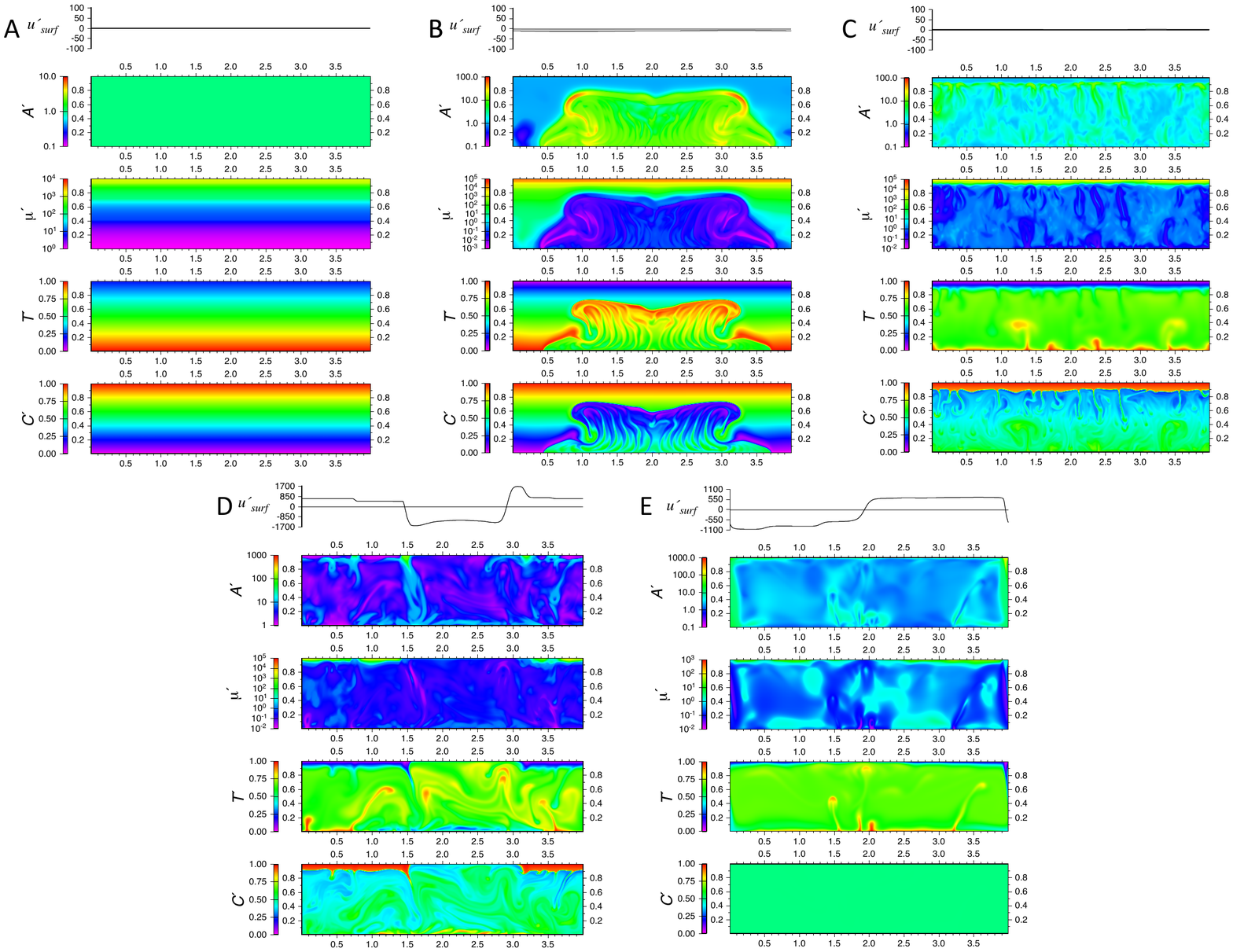}
\caption{\label {fig:init_pform} Images from a thermo-chemical model showing showing A) the initial condition, B) the initial overturn, C) post-overturn stagnant lid convection, D) the initiation of subduction, and E) the final statistical steady-state convection.  Each image shows (from top to bottom), the horizontal surface velocity ($u_{surf}'$), the fineness field ($A'$), the viscosity field ($\mu'$), the temperature field ($T'$), and the composition field ($C'$).  The model shown here uses $Ra_0 = 10^6$, $D = 10^{-2}$, $H = 10^3$, $E_v' = E_h' = 23.03$, $B=1$, and $Le=10^4$.}  
\end{figure}

\begin{figure}
\includegraphics[scale = 0.5]{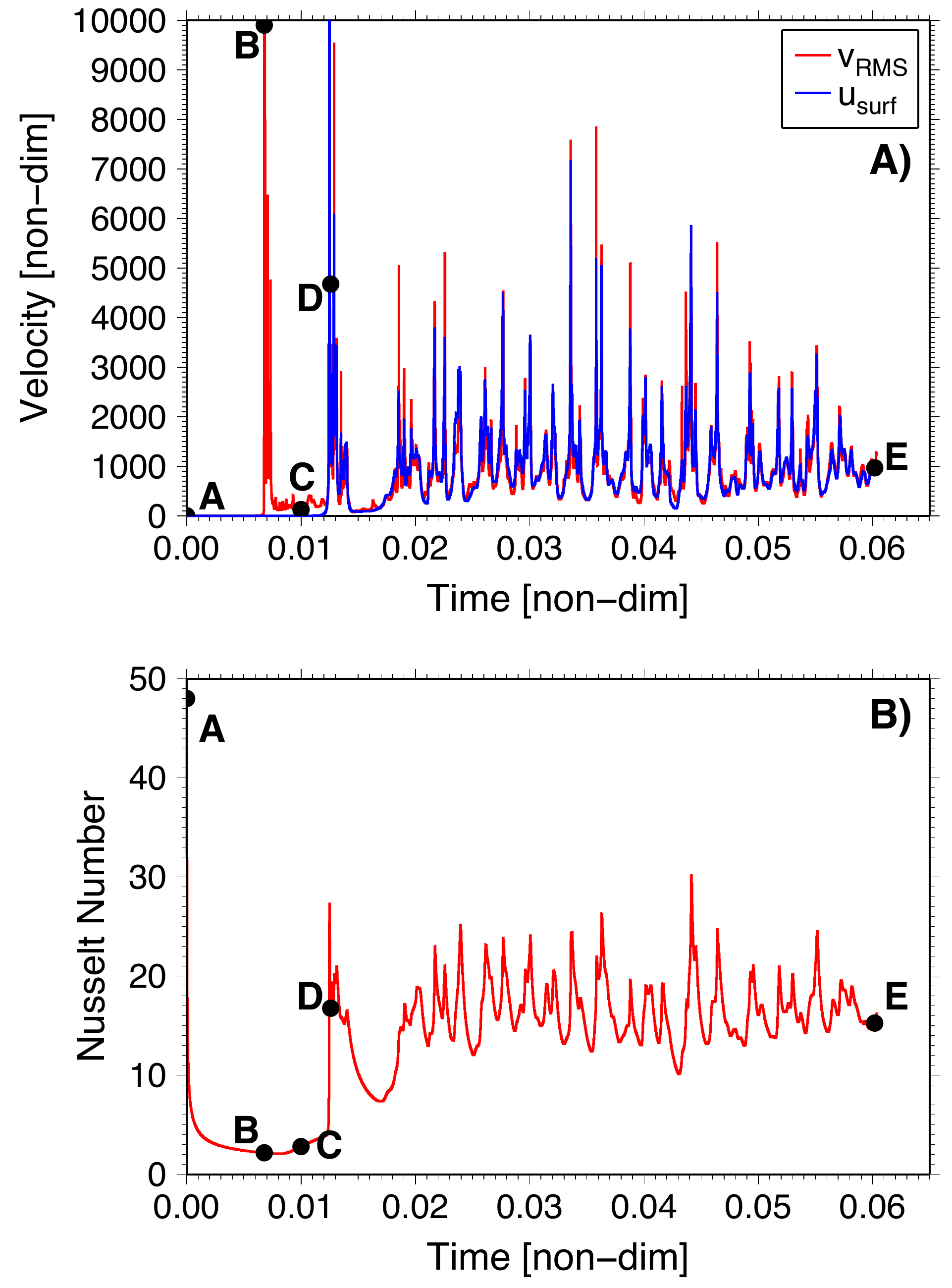}
\caption{\label {fig:vel_data} Velocity (A) and Nusselt number (B) time-series data for the numerical result shown in Figure \ref{fig:init_pform}.  Black dots labeled A-E mark the location of the images shown in panels A-E of Figure \ref{fig:init_pform}}  
\end{figure}

Most numerical simulations yield the same sequence of convective overturn followed by proto-plate-tectonic surface motion (Figures \ref{fig:init_pform} \& \ref{fig:vel_data}). First there is an initial Rayleigh-Taylor instability (Figure \ref{fig:init_pform}B), which takes place beneath a stagnant lid since the lithospheric viscosity is high and lithospheric damage is not sufficiently developed.  When compositional buoyancy is included, the sub-lid initial overturn leaves the chemically densest material in the rigid lid, while beneath the lid, the overturn results in a gravitationally stable compositional profile.  

A relatively short-lived period of stagnant lid convection takes place after the initial overturn (Figure \ref{fig:init_pform}C).  The stable chemical stratification following the first overturn does not preclude further convection since the upper boundary layer is cold and convectively unstable.  The ensuing thermal convection weakens the compositional heterogeneity in the mantle interior through stretching, mixing, and diffusing.  However, the compositionally dense material remains in the lithosphere during this phase of stagnant lid convection. With increasing $Ra_0$, the time spent in this period of stagnant lid convection shrinks, as the larger forcing of lithospheric damage, resulting from higher $Ra_0$, is able to initiate subduction more rapidly.    
         
During the period of stagnant lid convection, the lithosphere progressively thins and fineness increases in lithospheric shear zones.  These zones usually overlie drip-like downwellings emanating from the base of the rigid lid.  After sufficient damage has accumulated in the lithosphere, subduction initiates (Figure \ref{fig:init_pform}D). From here the calculation will generally continue in the mobile lid mode and observables such as plate speed and Nusselt number will settle to their expected values based on the scaling laws in \cite[e.g][]{Foley2013_scaling}.  Of course for the Earth, the plate speed and heat flow would continue to evolve as both mantle temperature, and possibly surface temperature, evolve.  However, as our numerical models are only meant to constrain the timescale for initiating mobile lid behavior, these effects of thermal evolution are excluded. Models with lower $Ra_0$ or larger $B$ have a brief period of sluggish convection after the first subduction event.  The initial period of subduction can happen so rapidly that the lithosphere is extensively thinned, and must grow diffusively in order to regain negative buoyancy.  However, lid mobility is not inhibited in this scenario, as inherited lithospheric weak zones allow subduction to occur once the lithosphere has regained sufficient negative buoyancy.  When chemical heterogeneity is included, the initiation of subduction flushes the chemically dense material from the lithosphere; this dense material then accumulates at the base of the mantle where subsequent convective motions mix and erase the heterogeneity.    

Even after settling into a mobile-lid regime, our models display a strong time-dependence in plate-speed and heat flow (Figure \ref{fig:vel_data}), stronger than is typically seen from pseudoplastic models in the plate-tectonic regime \citep[e.g.][]{Moresi1998,Tackley2000a,Korenaga2010}. The increased time-dependence in our models is likely a result of using grain-damage instead of pseudoplasticity, as grain-damage involves time-evolving lithospheric and mantle weakening, which can naturally introduce a stronger time-dependence in the convection calculations. However, our mobile lid results, whether in the sluggish subduction regime or in a plate tectonic-style regime, do not alternate between stagnant lid and mobile lid behavior; the lid remains mobilized throughout the time-dependent oscillations. Therefore, our mobile lid models are different from what is typically classified as episodic regime convection \citep[e.g.][]{Moresi1998}, though the exact boundary between episodic convection and plate-tectonic convection is somewhat arbitrary. Furthermore, the time-dependence observed in our numerical models is likely stronger than what would be expected on a planet, because mantle convection models with plate-generating rheologies often display stronger time-dependence in 2-D than in 3-D \citep[e.g.][]{Tackley2000a}.

\subsection{Overturn, Onset, and Lag Times}
\label{sec:results_times}

\begin{figure}
\includegraphics[scale = 0.6]{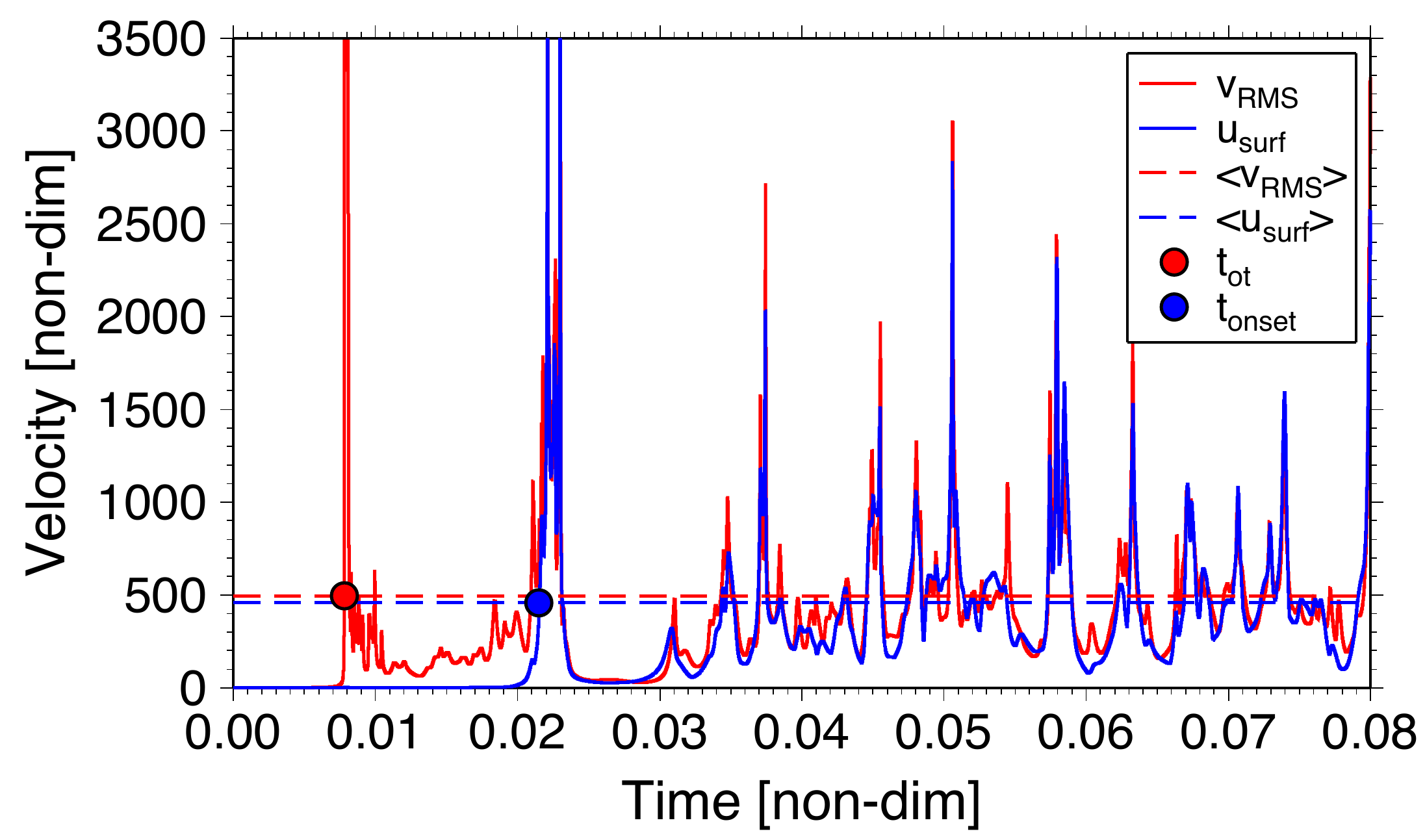}
\caption{\label {fig:vel_example} Plot of whole mantle rms velocity ($v'_{rms}$), surface velocity ($u'_{surf}$), time-averaged rms and surface velocity ($\langle v'_{rms} \rangle$ and $\langle u'_{surf} \rangle$), and the overturn time $t'_{ot}$ and onset time $t'_{onset}$.  The numerical result shown here uses $Ra_0 = 10^6$, $D = 2 \times 10^{-3}$, $H = 10^3$, $E_v' = E_h' = 23.03$, and $Le=10^4$.}  
\end{figure}

In order to understand the timing for the initiation of both convection and proto-plate tectonics, we determine the overturn time, the time at which the initial thermal or thermo-chemical overturn takes place, the onset time, the time of the first proto-subduction event, and the lag time, the time elapsed from the initial thermal or thermo-chemical overturn to the first subduction event.  We denote the overturn time as $t_{ot}$, the onset time as $t_{onset}$, and the lag time as $t_{lag} = t_{onset} - t_{ot}$. We define the overturn time as the time when the whole mantle rms velocity first reaches the time averaged rms velocity of the model run in statistical steady-state (Figure \ref{fig:vel_example}). Likewise, the onset time is defined as the time when the surface velocity first reaches the time averaged surface velocity (Figure \ref{fig:vel_example}).  From the numerical models, we report the non-dimensional overturn time, $t_{ot}'$, non-dimensional onset time, $t_{onset}'$, and non-dimensional lag time, $t_{lag}'$, where all quantities have been non-dimensionalized by $d^2/\kappa$. 

We measured the overturn, onset, and lag times for a large suite of numerical models where we vary the key parameters $D$, $Ra_0$, $\mu_l/\mu_m$ (via the activation energy $E_v'$), $h_l/h_m$ (via $E_h'$), $Q'$, and the initial perturbation amplitude; most sets of experiments include both cases with chemical heterogeneity where $B=1$, and those without chemical heterogeneity where $B=0$. We also ran thermochemical models where $Le$ and $B$ are varied to assess the influence of these parameters.           

\begin{figure}
\includegraphics[scale = 0.5]{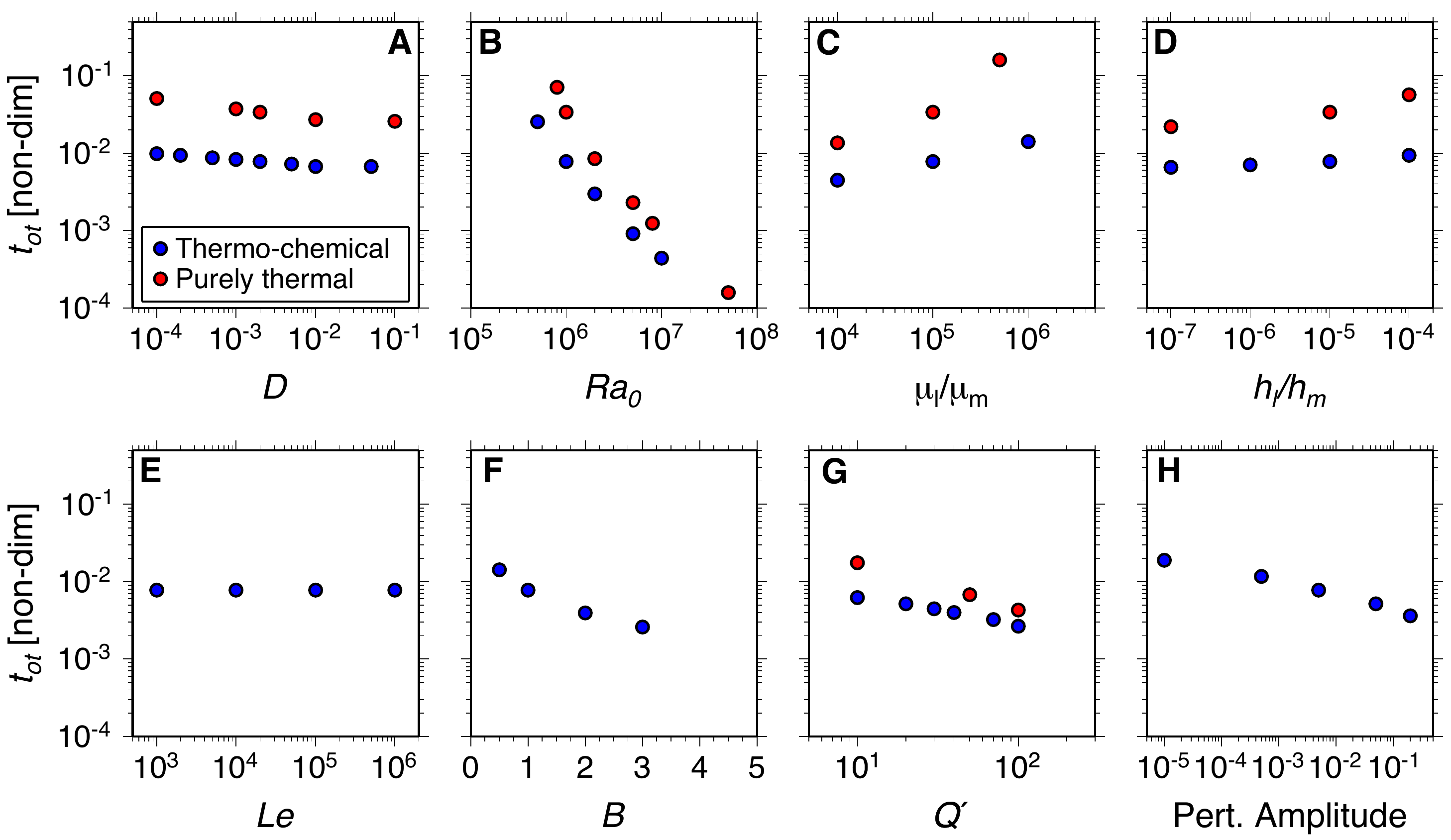}
\caption{\label {fig:overturn_time} Overturn time versus $D$ (A), $Ra_0$ (B), $\mu_l/\mu_m$ (C), $h_l/h_m$ (D), $Le$ (E), $B$ (F), $Q'$ (G), and initial perturbation amplitude (H).  All the models use $m=2$, $p=4$, and thermo-chemical models use $B=1$ and $Le=10^4$, except in (E) and (F) where those parameters are varied.  Models in (G) are purely internally heated, while the rest of the models are bottom heated with $Q' = 0$.  All models have an initial perturbation amplitude of $5 \times 10^{-3}$, except for those in (H) where the amplitude is varied.  The rest of the parameters are (A) $H=10^3$, $Ra_0=10^6$, $E_v'=E_h'=23.03$ (resulting in $\mu_l/\mu_m = 10^5$ and $h_l/h_m = 10^{-5}$); (B) $H=10^3$, $D = 2 \times 10^{-3}$, $E_v'=E_h'=23.03$; (C) $H=10^3$, $D = 2 \times 10^{-3}$, $Ra_0 = 10^6$, $E_h'=23.03$; (D) $H=10^3$, $D = 2 \times 10^{-3}$, $Ra_0 = 10^6$, $E_v'=23.03$; (E), (F), (G), and (H) $D = 2 \times 10^{-3}$, $H=10^3$, $Ra_0 = 10^6$, $E_v'=E_h'=23.03$. }  
\end{figure}
         
The overturn time decreases with either increasing $Ra_0$ or increasing $D$ (Figure \ref{fig:overturn_time}A,B), because a larger Rayleigh number means stronger relative buoyancy forces, and a larger damage number causes more viscosity reduction in the mantle as instabilities develop, hastening the overturn. However, the influence of $D$ on $t_{ot}'$ is significantly less than the influence of $Ra_0$.  Increasing $\mu_l/\mu_m$ (via a larger activation energy) causes a longer $t_{ot}'$ (Figure \ref{fig:overturn_time}C), by increasing the mantle viscosity and thereby retarding the growth of instabilities. Increasing $h_l/h_m$ (via the healing activation energy) increases the overturn time, but only by a small amount, similar to the effect of $D$ (Figure \ref{fig:overturn_time}D). Varying $Le$ has almost no effect on $t_{ot}'$, indicating that our primary choice of $Le$ does not impact the results (Figure \ref{fig:overturn_time}E), while larger $B$ leads to shorter overturn times because there is increased buoyancy in the mantle driving the onset of convection (Figure \ref{fig:overturn_time}F).  Finally, increasing either the internal heating rate, $Q'$, or the initial perturbation amplitude decreases the overturn time (Figures \ref{fig:overturn_time}G,H).  Larger internal heating rates warm up the mantle, reducing the viscosity and allowing perturbations to grow more rapidly, while a larger initial perturbation will naturally reach finite amplitude more quickly.

\begin{figure}
\includegraphics[scale = 0.5]{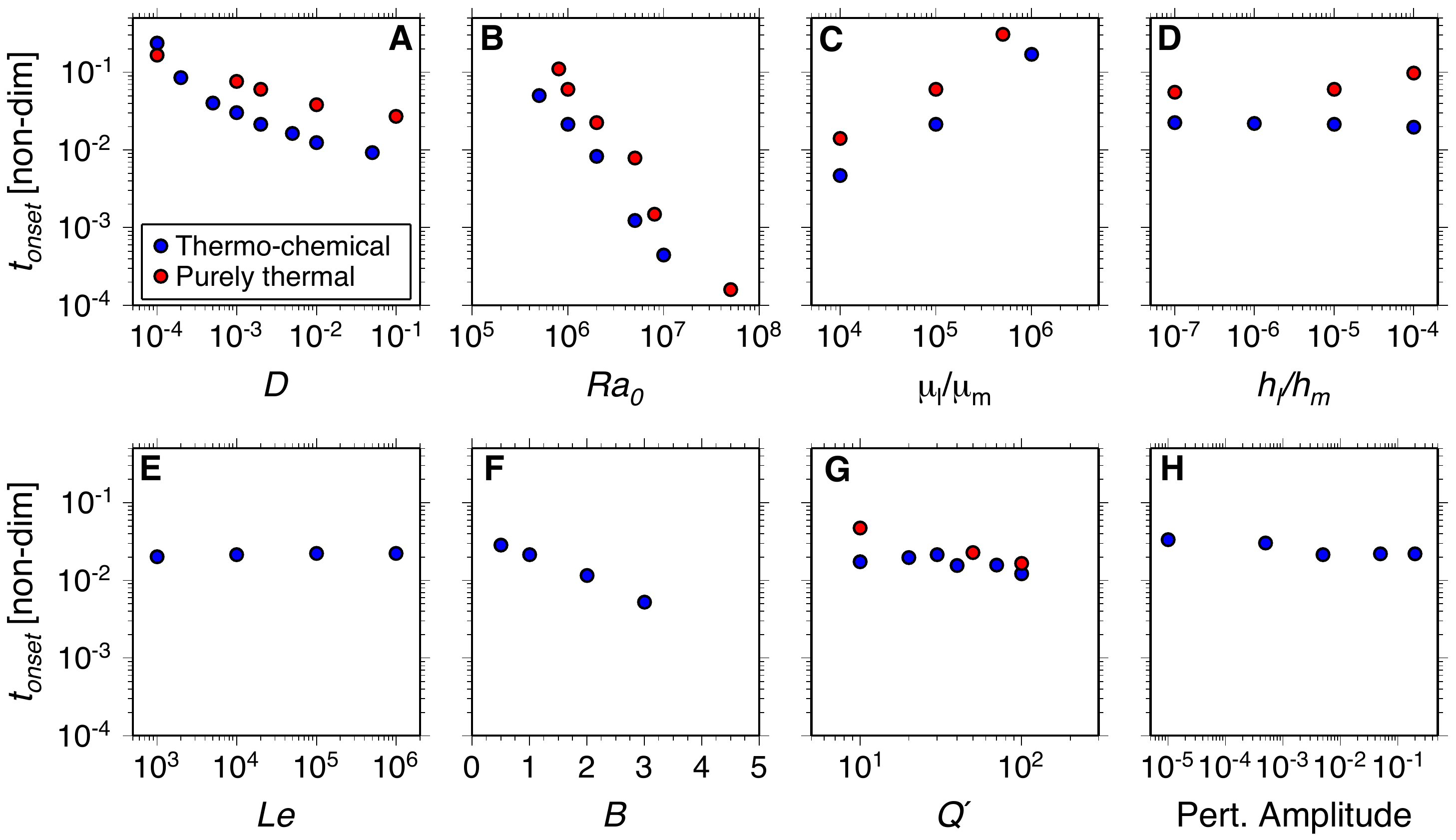}
\caption{\label {fig:onset_time} Onset time versus $D$ (A), $Ra_0$ (B), $\mu_l/\mu_m$ (C), $h_l/h_m$ (D), $Le$ (E), $B$ (F), $Q'$ (G), and initial perturbation amplitude (H). Parameters are listed in the caption to Figure \ref{fig:overturn_time}.}  
\end{figure}

\begin{figure}
\includegraphics[scale = 0.5]{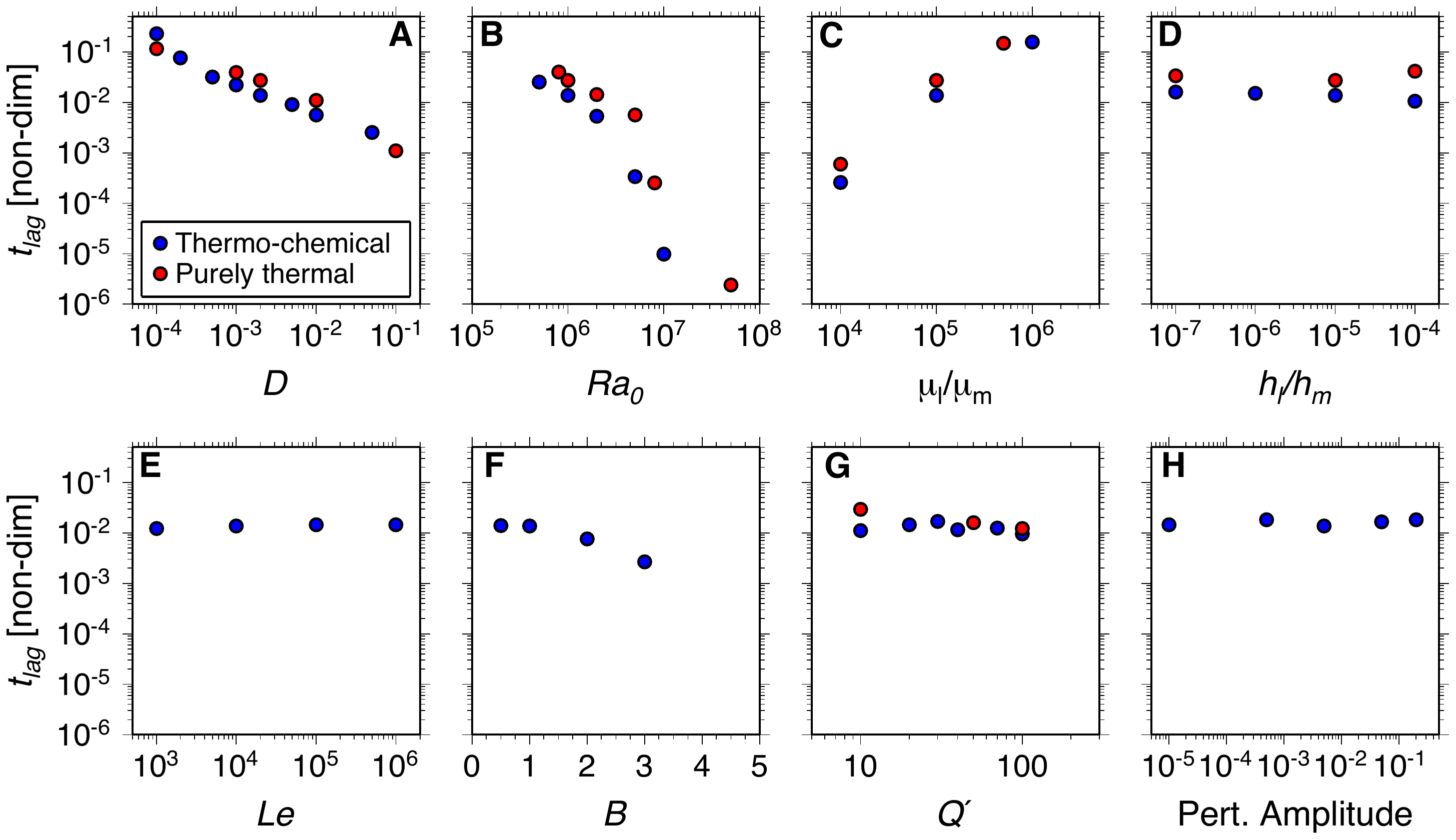}
\caption{\label {fig:lag_time} Lag time versus $D$ (A), $Ra_0$ (B), $\mu_l/\mu_m$ (C), $h_l/h_m$ (D), $Le$ (E), $B$ (F), $Q'$ (G), and initial perturbation amplitude (H). Parameters are listed in the caption to Figure \ref{fig:overturn_time}.}  
\end{figure}
      
The numerical data for the lag time and the onset time both show some clear trends (Figures \ref{fig:onset_time} \& \ref{fig:lag_time}).  The onset time, measuring the time elapsed from the start of the simulation until subduction initiates, essentially includes the overturn time and is therefore influenced by the initial perturbation.  We thus focus on the lag time results as the lag time subtracts out the contribution from the overturn time, and as a result removes the influence of the initial perturbation. Nonetheless, the trends discussed for the lag time are broadly consistent with those for the onset time.  As before, increasing $Ra_0$ or $D$ both cause shorter lag times (Figure \ref{fig:lag_time}A,B), because subduction initiates after sufficient damage has built up in lithospheric shear zones, which happens sooner with given greater convective vigor and/or more efficient damage.  Furthermore, the decrease in $t_{lag}'$ with $Ra_0$ is non-linear; the magnitude of the slope (in log-log space) increases as $Ra_0$ increases, meaning that the lag time collapses towards 0 at high $Ra_0$.  At very large Rayleigh numbers mobile lid convection is initiated almost concurrently with the beginning of convection.  Larger viscosity ratios, $\mu_l/\mu_m$, increase $t_{lag}' $ (Figure \ref{fig:lag_time}C), because more damage must accumulate in a relatively more viscous lithosphere before proto-subduction can initiate.  Varying the healing ratio, $h_l/h_m$, has almost no impact on $t_{lag}'$; i.e. slower grain-growth in the lithosphere does not help proto-subduction initiate more rapidly, nor does faster grain-growth slow initiation (Figure \ref{fig:lag_time}D). As plate initiation is determined by how quickly damage can build up in the lithosphere, it is thus dominated by the grain-reduction term, and the effects of grain-growth are minor (grain-growth is, however, important for determining the final steady state to which the system evolves after initiation). Consistent with the $t_{ot}'$ results, the presence of chemical heterogeneity almost always results in a shorter $t_{lag}'$, because the presence of chemically dense material in the lithosphere after the initial overturn provides more buoyancy stress to drive damage in lithospheric shear zones.  Increasing $B$ causes a decrease in $t_{lag}'$ for this same reason: there is more negative buoyancy available in the lithosphere to drive damage and subduction (Figure \ref{fig:lag_time}E).  The Lewis number has almost no influence on the onset time, confirming that the level of chemical diffusion in our models does not significantly impact the results (Figure \ref{fig:lag_time}F). Finally, the lag time is nearly insensitive to internal heating rate and the amplitude of the initial perturbation (Figure \ref{fig:lag_time}G,H).  Higher internal heating rates increase the effective Rayleigh number of the mantle, by increasing the temperature difference driving convection and reducing the mantle viscosity; however, the effective viscosity ratio also increases, roughly canceling out the influence of a higher Rayleigh number.  The insensitivity of $t_{lag}'$ to the initial perturbation amplitude indicates that the timescale for initiating mobile lid behavior, once convection has begun, is not sensitive to the details of the initial condition.         

\section{Scaling Analysis} 
\label{sec:scaling}

In order to produce a meaningful estimate for the initiation time of proto-plate tectonics in the early Earth, and to more fully understand the governing physics behind the initiation of proto-plate tectonics, we determine a scaling law for $t_{lag}'$.  As the overturn time is an important aspect of the overall evolution to proto-subduction, we first briefly discuss how grain-damage influences $t_{ot}'$.  

\subsection{Overturn Time}
\label{sec:overturn_time2}

\begin{figure}
\includegraphics[scale = 0.5]{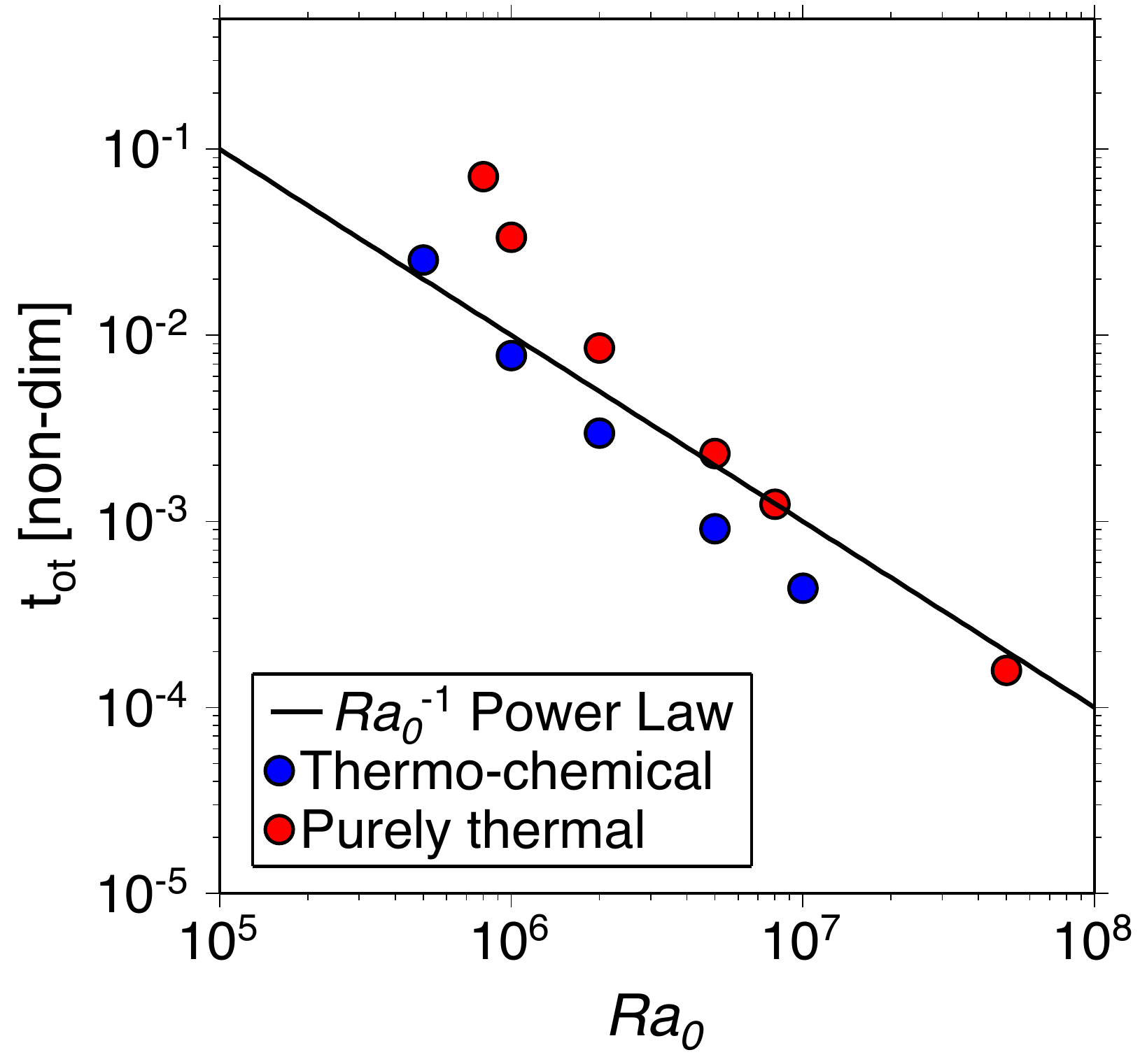}
\caption{\label {fig:ot_2} Plot of the overturn time versus $Ra_0$ with an $Ra_0^{-1}$ power law.}  
\end{figure}

The initial overturn is essentially a Rayleigh-Taylor instability.  Numerical results approximately follow the classical scaling law for the timescale of a Rayleigh-Taylor instability, $t' \sim Ra_0^{-1}$ (Figure \ref{fig:ot_2}) \citep[e.g.][]{turc1982}. This result, and the observed weak influence of the damage parameters (e.g. $D$, $h_l/h_m$) on $t_{ot}'$, indicate that the initial Rayleigh-Taylor instability is predominately governed by the buoyancy of the system.  The initial overturn occurs rapidly, not giving damage sufficient time to build up in the mantle during the overturn; thus the effects of damage are muted for this event.

\subsection{Lag Time}

We develop an empirical scaling law for the lag time as a function of the various governing parameters for thermo-chemical convection with grain-damage.  As shown by the numerical results (\S \ref{sec:results_times}), $t'_{lag}$ is nearly independent of healing because the initiation of plate tectonics is determined by how quickly damage can build up in the lithosphere, and is thus dominated by the grain-reduction term.  The lag time is also found to be nearly insensitive to internal heating rate, Lewis number, and initial perturbation amplitude.  We therefore assume that $t'_{lag}$ is solely a function of damage number, Rayleigh number, chemical buoyancy, and viscosity ratio, and determine empirical fits to the numerical data varying these parameters.

\begin{figure}
\includegraphics[scale = 0.8]{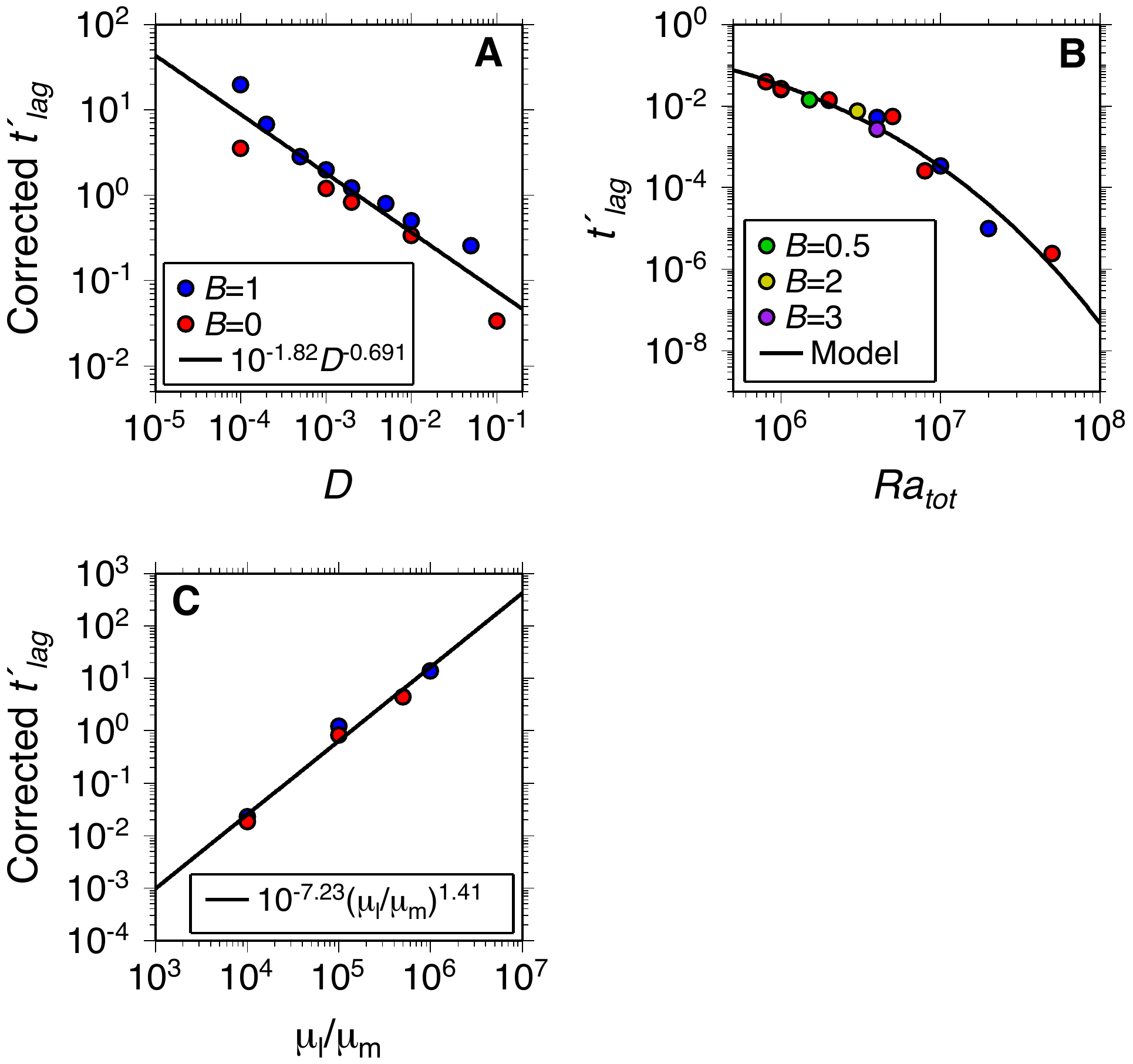}
\caption{\label{fig:fit_onset} Empirical fits to the numerical results for $t'_{lag}$ as a function of $D$ (A), $Ra_{tot}$ (B), and $\mu_l/\mu_m$ (C).  The empirical fits to the corrected lag time (corrected lag time is defined in the text below equation \eqref{onset_time}) in (A) and (C) are to simple power law functions, with the results listed in the figure legend, while the fit in (B) is to equation \eqref{Ra_tot_fit}. Parameters for the numerical model results displayed here are listed in the caption to Figure \ref{fig:overturn_time}.  }
\end{figure}

We next assume that the effects of chemical buoyancy can be incorporated into the scaling law by defining a total Rayleigh number, $Ra_{tot} = \rho g \alpha \Delta T d^3 (1+B)/(\kappa \mu_m)$, that includes both thermal and chemical buoyancy.  This approach is able to collapse all the data, with $B$  ranging from 0 to 3, onto one curve in $t'_{lag} - Ra_{tot}$ space (Figure \ref{fig:fit_onset}B).  The $t_{lag}' - D$ and $t_{lag}' - \mu_l/\mu_m$ trends are linear in log-log space, indicating that $t_{lag}'$ has a simple power law dependence on these variables.  However, the dependence of $t_{lag}'$ on $Ra_{tot}$ is non-linear, and thus the functional dependence between these two variables is more complicated.  The simplest equation that can fit the numerical results with varying $Ra_{tot}$ is 

\begin{equation}
\eqlbl{Ra_tot_fit}
\log_{10} (t_{lag}') =  a_1 \log_{10} (Ra_{tot})^{a_2} .
\end{equation}
Performing a least-squares fit to the numerical results where $Ra_{tot}$ is varied, we find that $a_1 = -7.2 \times 10^{-5}$ and $a_2 = 5.545$ (Figure \ref{fig:fit_onset}B).  Combining \eqref{Ra_tot_fit} with the assumed power law dependencies for $D$ and $\mu_l/\mu_m$, the final scaling law for the lag time is   

\begin{equation} 
\eqlbl{onset_time}
t'_{lag} = 10^{a_1 \log_{10}(Ra_{tot})^{a_2}} \left(\frac{D}{D^*}\right)^{\beta_D} \left(\frac{(\mu_l/\mu_m)}{(\mu_l/\mu_m)^*}\right)^{\beta_{\mu}} 
\end{equation} 
where $\beta_D$ and $\beta_{\mu}$ are the scaling exponents, and $D^* = 2 \times 10^{-3}$ and $(\mu_l/\mu_m)^* = 10^5$ are the values of these parameters used for the numerical experiments with varying $Ra_{tot}$.  We use a least-squares fit to the corrected lag times (where corrected lag time is $t_{lag}'/10^{a_1 \log_{10}(Ra_{tot})^{a_2}}$) for the numerical experiments with varying $D$ and $\mu_l/\mu_m$ and find $\beta_D \approx -0.691$ and $\beta_{\mu} \approx 1.41$ (Figure \ref{fig:fit_onset}). 

\section{Application of Lag Time Scaling Law to the Hadean Earth}
\label{sec:app}

\begin{figure}
\includegraphics[scale = 0.6]{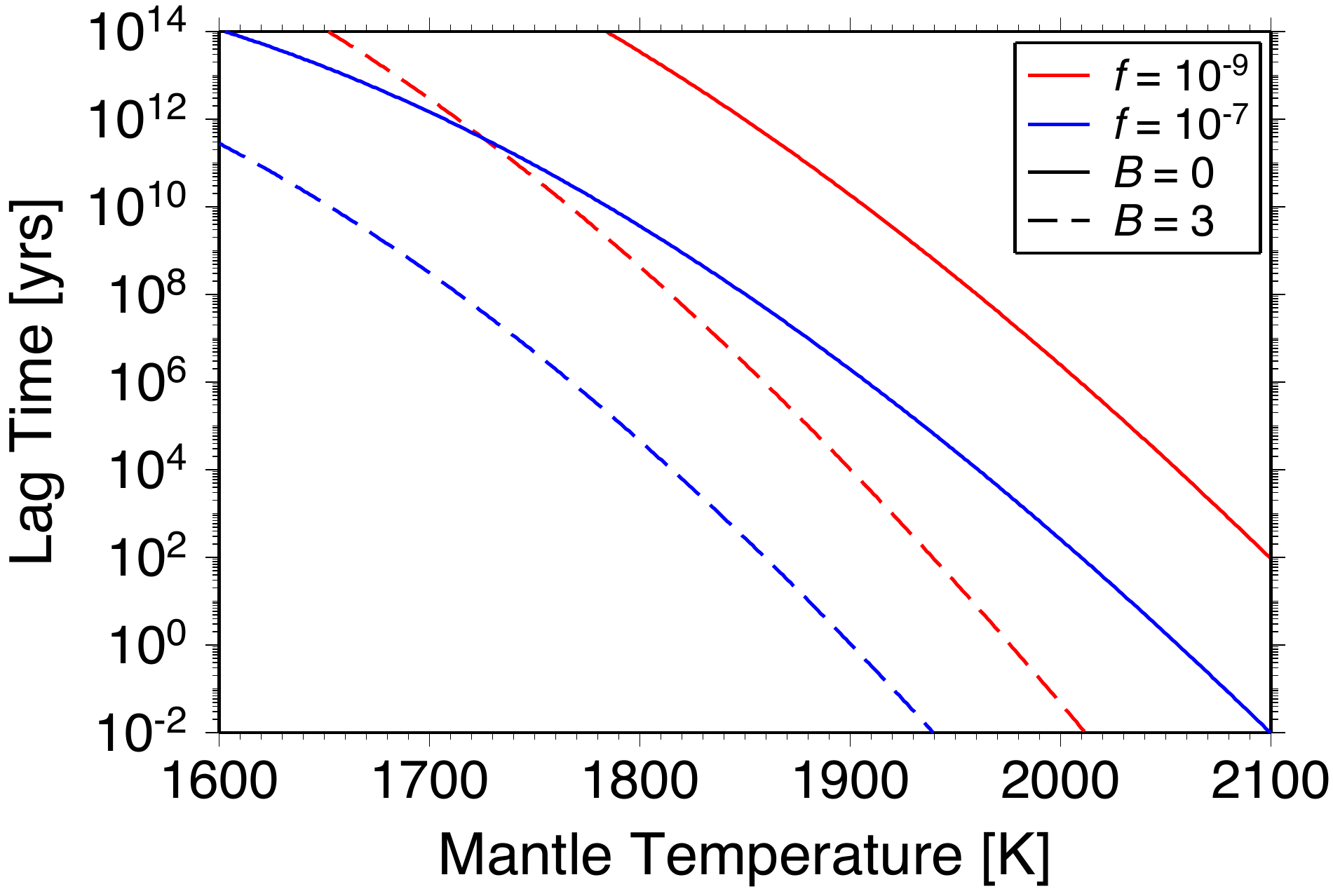}
\caption{\label {fig:plate_initiate} Plot of the lag time versus interior mantle potential temperature, at two different values of $f$: $f = 10^{-9}$ (red line) and $f = 10^{-7}$ (blue line).  For each value of $f$, curves are plotted with $B=0$ (solid line), and $B=3$ (dashed line).}  
\end{figure}

We use our scaling law for the lag time \eqref{onset_time} to place constraints on how rapidly proto-subduction could begin after magma ocean solidification.  As in \S \ref{sec:hadean_scale}, we estimate the lag time, the time from when convection begins to when proto-subduction initiates, as a function of mantle temperature, scaling up to Hadean conditions.  We also assume that the scaling law for $t_{lag}$ \eqref{onset_time} can be applied to the internally heated Earth because 1) our numerical results are largely insensitive to internal heating rate and heating mode, and 2) the internal temperatures in our bottom heated numerical models (after convection has begun), are approximately equal to the basal mantle temperature, and thus $Ra_0 \approx Ra_i$, $\mu_m \approx \mu_i$, and $D \approx D \mu_i/\mu_m$ as in \S \ref{sec:scale_model}. As a result, the mantle temperatures discussed here are again interior mantle potential temperatures.  We use the same parameters from \S \ref{sec:Tl_param} for the damage number, lithospheric healing rate, and internal Rayleigh number, and calculate the lithosphere temperature in the same manner, i.e. using \eqref{lith_temp}.  We use two different values of $f$, to test the influence of varying the efficacy of damage: $f \approx 10^{-9}$ (the value of $f$ used in \S \ref{sec:Tl_param}) and $f \approx 10^{-7}$ which is closer to estimates from laboratory and field studies \citep{austin2007,Rozel2010}.  Finally, for each value of $f$, we compute the lag time for $B=0$ and $B=3$, i.e. the range of plausible $B$ for the Hadean Earth, to show the influence chemical heterogeneity has on $t_{lag}$. We also assume $T_s=500$ K, to test if proto-plate tectonics can initiate in a timely manner in the presence of a CO$_2$ rich primordial climate. 

Increasing mantle temperature sharply decreases the lag time (Figure \ref{fig:plate_initiate}), because a warmer mantle has a higher Rayleigh number, and a higher Rayleigh number leads to stronger driving forces for lithospheric damage.  At the lowest value of $f$, the lag time is $\approx 100$ Myrs at $T_i \approx 2000$ K, and decreases to $\approx 100$ yrs, effectively initiating proto-subduction simultaneously with the beginning of solid-state mantle convection, at $T_i \approx 2100$ K.  Varying $f$ also has a large impact on $t_{lag}$; increasing $f$ by 2 orders of magnitude results in a decrease in $t_{lag}$ of approximately 4 orders of magnitude.  The value of $f \approx 10^{-9}$, as used in \S \ref{sec:Tl_param}, can be considered a minimum value, as experimental estimates are larger.  Thus, uncertainties in $f$ should only act to decrease the lag time.  Finally, including chemical buoyancy also drops the lag time significantly by boosting the effective Rayleigh number of the system. 

Our results show that at a post-magma ocean mantle temperature of $2100$ K, the lag time is extremely short, effectively zero on a geologic timescale, even at the minimum value of $f$ and without chemical buoyancy.  The lag time increases to $\approx 100$ Myrs at $T_i \approx 2000$ K, climbing even higher at lower temperatures, again for the lowest value of $f$ and with $B=0$; however the higher mantle temperatures are probably most applicable to the Hadean Earth.  Thus, our results indicate that proto-subduction can initiate rapidly after solid-state mantle convection begins.  We complete our analysis of the timescale for initiating proto-subduction by estimating the overturn time for the onset of convection from a simple scaling law describing the development of a Rayleigh-Taylor instability; we neglect the effects of grain-damage in this analysis because our numerical models showed they are unimportant for the initial overturn (see \S \ref{sec:overturn_time2}).  The overturn time can be approximated as the time required for a sphere, of radius $R_s$, to rise across the mantle, and thus $t_{ot} \approx d/v_s$, where $v_s$ is the Stoke's velocity of the sphere.  Using the expression for the Stoke's velocity from \cite{Batchelor1967}, the overturn time (in seconds) is,  

\begin{equation}
\eqlbl{RT}
t_{ot} \approx \left(\frac{9}{2}\right) \frac{\mu_i d }{\Delta \rho g R_s^2} ,
\end{equation} 
where $\Delta \rho$ is the average density difference between the buoyant sphere and the background mantle, $\Delta \rho = (\rho \alpha \Delta T(1+B))/2$; for the most conservative estimate, we assume $B=0$ and find $\Delta \rho \approx 160$ kgm$^{-3}$ for $\Delta T \approx 2700$ K \citep[e.g.][]{Herzberg1996,solomatov2000}.  The spherical upwelling representing Rayleigh-Taylor instability will have a radius that is some fraction of the mantle thickness; we conservatively assume $R_s = 30-300$ km, or 1 \% to 10 \% of the mantle thickness (larger radii will give shorter overturn times). With a viscosity of $\mu_i \approx 10^{19}$ Pa s, appropriate for the mantle near its solidus, we find a range of $t_{ot} \sim 10^4$ yrs for $R_s = 300$ km and $t_{ot} \sim 10^6$ yrs for $R_s = 30$ km. Combining the estimate for the overturn time with the estimate for the lag time, our model demonstrates that proto-subduction could initiate within 100 Myrs or less for mantle temperatures 2000 K or larger, which are expected for the Hadean Earth. Thus the operation of proto-subduction on the very early Earth was geodynamically plausible. 

\section{Discussion}
\label{sec:discussion}

Our results predict proto-subduction on the very early Earth, likely within $\sim 100$ Myrs of magma ocean solidification, based on analysis from scaling laws that predict the ``equilibrium," or statistically steady-state mode of convection, and numerical models that demonstrate this ``equilibrium" mode can be reached rapidly.  The effects of high interior mantle and high surface temperatures in the Hadean are not sufficient to eliminate lithospheric recycling (\S \ref{sec:scale_results}), and the consequences of damage and vigorous early Earth mantle convection allow proto-subduction to initiate during the Hadean (\S \ref{sec:app}).  Furthermore, proto-subduction was distinct from modern day plate tectonics, being characterized by drip-like downwellings and diffuse plate boundaries.  Our results are consistent with geological observations of early Earth subduction.  Hadean zircons provide evidence for crustal recycling \citep[e.g.][]{Watson2005} and low heat flows \citep{Hopkins2008}, both possible indicators of subduction.  Furthermore, the appearance of these indicators by 4.4-4.3 Ga implies a $\sim 100$ Myr onset time for Hadean proto-subduction, also consistent with our results.  However, the interpretation of Hadean zircon data is still highly uncertain, so the possibility of a later onset time for proto-plate tectonics can not be ruled out.  Finally, our scaling law analysis indicates that early Earth proto-subduction, as inferred from Hadean zircon data, was different from modern style plate tectonics, potentially explaining why Hadean proto-subduction seems to lack some of the petrological, geochemical, and tectonic features of modern day plate tectonics \citep[e.g.][]{Condie2008,Shirey2011,Debaille2013}.  The interior mantle temperature, and possibly the surface temperature, may need to cool before modern day plate tectonics can develop.

Based on our numerical and scaling law results, the timeline of events after proto-plate initiation is as follows. First, the initial subduction event occurs as a rapid foundering of the entire lithosphere, with plate speeds significantly higher than what would be predicted from scaling laws for convection in statistical steady-state (e.g. \S \ref{sec:scale_results}).  The lithosphere is thinned extensively after the initial phase of subduction, and this may result in a brief period of sluggish convection as the new boundary layer grows diffusively until it becomes convectively unstable.  However, the presence of damaged shear zones inherited from the initial subduction event allow a mobile surface mode of convection to persist.  Convection then converges to the statistically steady-state behavior predicted by the scaling laws for convection with grain-damage after a few (i.e. $\sim 2-3$) overturns. However, proto-subduction is not episodic during this transitory phase, as we do not observe periods of fully fledged stagnant lid behavior between overturns in our numerical models. Both the initial subduction event, and the ensuing mobile lid convection, will take place in a sluggish subduction regime, where plate boundaries are more diffuse than observed for modern day plate tectonics, due to the effects of grain-growth in the lithosphere.  However, as seen in the scaling law results of \S \ref{sec:scale_results}, as the mantle and surface temperatures cool, plate speed increases and the modern style of plate tectonics develops.  In fact, early Earth proto-subduction could significantly aid in the development of modern style plate tectonics by drawing a large quantity of CO$_2$ out of the post-magma ocean atmosphere. However, determining the exact timing of these events after the initiation of proto-plate tectonics would require a detailed thermal history model, and is left for future studies.  

In particular, Hadean proto-subduction could be important for the development of clement surface conditions, because recycling of surficial carbon into the mantle may be necessary to cool Earth's early climate.  \cite{Sleep2001a} shows that complete carbonization of the upper 500 m of oceanic crust, the depth to which oceanic crustal carbonization proceeds to on the Earth today, can only remove a small fraction of the carbon present in a CO$_2$ rich proto-atmosphere; thus subduction of carbon into the mantle is necessary to cool the climate.  Proto-subduction would allow sluggish crustal recycling into the mantle from very early in Earth's history, and would therefore help remove any possible excess atmospheric CO$_2$, with the caveat that the petrology of carbon carriage into the mantle via subduction is a complex, and incompletely understood process \citep[e.g.][]{Ague2014}.  Future studies could add our model for the evolution of Hadean tectonics into simple carbon cycle box models \citep[e.g.][]{Driscoll2013}, to constrain the timescales of carbon cycling between the primordial atmosphere and mantle, and the implications this has for the subsequent evolution of the Earth.     

Convection in the Hadean likely produced a significant amount of melting, and this may influence the lag time and the operation of proto-plate tectonics.  In fact, melting and volcanism has been proposed as the primary mantle heat loss mechanism for the early Earth, both in models that find plate tectonics is not in operation \citep{Moore2013}, and in those that have mobile plates \citep{Nakagawa2012}.  One possible effect of large scale upper mantle melting would be the production of fractures and damaged zones where melt migrates to the surface and erupts.  These damaged zones could then serve as nucleation points for lithospheric shear zones, allowing plate tectonics to initiate faster than in the initially pristine lithosphere we considered. Melting could also produce a thick buoyant crust; however this crust would lie on top of a thick, cold, and possibly chemically dense lithosphere, so there would have to be a substantial melt layer to counteract the negative buoyancy of the proto-lithosphere. Nevertheless, understanding how the effects of melting influence the onset time and the operation of proto-subduction is an important avenue for future studies. 

\subsection{Comparison to Previous Studies}
\subsubsection{Hadean Plate Tectonics}
\label{sec:comp_previous}

Few other studies have considered how the rheological effects necessary for plate tectonics might change on the early Earth.  Using a pseudoplastic yield stress rheology, \cite{ONeill2007b} argued that the hotter mantle temperature of the early Earth would decrease lithospheric stresses, and could push the Archean Earth into a stagnant lid or episodic mode of convection. However, using new scaling laws developed from numerical models, \cite{Korenaga2011} argues that if the mantle is initially dry, a possible result of degassing during magma ocean solidification, plate tectonics could operate throughout Earth's history because the higher mantle viscosity would increase convective stresses. We find that with grain-damage, persistent plate motion and subduction could occur in the Hadean, similar to the result of \cite{Korenaga2011}, though dehydration of the primordial mantle is not necessary in our model.  The reason we do not find a transition to episodic or stagnant lid convection at high mantle temperatures, as in \cite{ONeill2007b}, or require a dry, high viscosity mantle to avoid this fate, as in \cite{Korenaga2011}, comes down to the fact that the driving forces for lithospheric weakening with grain-damage are different than those for pseudoplasticity, as we discuss further in the next paragraph.  Moreover, we also find that the style of convection in the Hadean was most likely not truly plate-tectonic, because less efficient lithospheric damage would lead to a more drip-like style of subduction, with diffuse plate boundaries. 

Similar to \cite{ONeill2007b}, many convection studies with the pseudoplastic rheology have found that high rates of internal heating, leading to high interior mantle temperatures, can cause a transition to episodic or stagnant lid convection \citep[e.g.][]{stein2004,Stein2013,Moore2013}.  The physical interpretation of this behavior is that the lower mantle viscosity decreases the typical stress scales in the mantle and lithosphere, and convection is no longer able to exceed the lithosphere's yield strength.  We find that increasing the internal heating rate does not cause a transition to stagnant lid behavior with grain-damage (see Figure \ref{fig:lag_time}, where mobile lid convection is maintained, with similar lag times, as internal heating rate increases), which reveals a key difference between pseudoplasticity and grain-damage.  The ability of pseudoplasticity to generate mobile lid convection is highly sensitive to the state of stress in the lithosphere and mantle, because the stress state directly determines whether yielding occurs or not.  Grain-damage is driven by the deformational work in the lithosphere, rather than depending only on the stress, and is therefore much less sensitive to stress variations caused by changes to internal temperature or viscosity.  For example, a decrease in mantle viscosity may drop the convective stress, but it will also increase the strain rate in the lithosphere.  We find that the primary role mantle temperature plays in plate generation with grain-damage is through its influence on lithospheric grain-growth rates (\S \ref{sec:scale_results}).  A similar argument explains why our models do not display the same hysteresis observed by \cite{Weller2012}; they find it is harder to develop mobile lid convection when the initial condition is in the stagnant lid regime, because the higher mantle temperature of the stagnant lid model drops the convective stress (the thicker lithosphere can also increase the effective yield strength).  Our models with grain-damage are not as sensitive to the convective stress, and therefore do not display the same hysteresis. 

Our finding that plate speed decreases with mantle temperature is also similar to the model of \cite{Korenaga2011}, where plate speed decreases with increasing mantle temperature because enhanced upper mantle melting causes dehydration stiffening of the lithosphere. Thus, our results show that the influence of mantle temperature on lithospheric healing provides an additional mechanism capable of producing sluggish early plate motions, independent of the mantle melting and dehydration stiffening model. That plate speed can decrease with increasing mantle temperature may also be important for solving the problem of crustal buoyancy preventing subduction (discussed in \S \ref{sec:intro}) because if plates move more slowly in the past, they can then grow thicker and accumulate more negative thermal buoyancy. The scaling laws of \cite{korenaga2006} and \cite{Korenaga2011}, where plate speed decreases with mantle temperature due to melt induced dehydration stiffening of the lithosphere, indicate that plates are still able to reach negative buoyancy even with a thick, buoyant crust, because plates subduct at an older age.  Grain-damage produces a similar scaling law, and therefore may be able to solve the buoyancy problem in the same manner. However, the effect of crustal buoyancy needs to be tested in numerical convection models to assess whether the predictions from scaling laws are borne out.  

Finally, numerical subduction models find that when mantle temperatures are increased, slabs tend to break-off and a more drip-like, intermittent style of subduction develops \citep{Vanhunen2008,Sizola2010,Vanhunen2012}.  The non-plate tectonic style of subduction observed in these models is qualitatively similar to the ``sluggish subduction" style of intermittent, drip-like lithospheric downwellings we predict for the early Earth.  Thus early Earth proto-subduction may have differed significantly from modern day plate tectonics, and understanding these differences could be vital for interpreting geological observations from the Archean and Hadean \citep[e.g.][]{Moyen2012}.  However, whether intermittent or drip-like subduction continues in these subduction models when mantle temperatures are increased all the way to Eoarchean or Hadean values depends on a number of factors, including slab strength, kinetics of the basalt to eclogite phase transition, and slab weakening due to asthenospheric melting. \cite{Sizola2010} finds that rapid slab break-off shuts subduction down when mantle temperatures are $250$ K warmer than the present day, because asthenospheric melting significantly weakens the slab, allowing it to detach.  In contrast, most of the models in \cite{Vanhunen2008} are able to sustain intermittent lithospheric dripping at mantle temperatures 300 K warmer than today, albeit without the effects of melt induced weakening modeled in \cite{Sizola2010}. In our models slab break-off does not completely shut down subduction either; our models are able to sustain drip-like proto-subduction, even at Hadean conditions, because grain-damage is still capable of providing a sufficient level of lithospheric weakening for the top thermal boundary layer to sink. 

\subsubsection{Initiation of Plate Tectonics}

The initiation of plate tectonics (or proto-plate tectonics) is a poorly studied problem.  There has been extensive work on initiating subduction, a similar problem, where mechanisms such as sediment loading paired with a wet rheology \citep{Rege2001} or weak zone reactivation \citep{gurnis2000} have been proposed.  However it is not clear if these mechanisms would be applicable to the early Earth.  Our study demonstrates that a plate generation mechanism, in our case grain-damage, is capable of explaining the initiation of proto-plate tectonics.  No special mechanism needs to be invoked; the same physics required for convection to occur in a plate-like manner can also explain the initiation of proto-subduction. There may be additional mechanisms, such as hydraulic fracturing and damage caused by melt migration, that could help initiate proto-plate tectonics more rapidly on the early Earth, but these are not required.  There are very few quantitative studies of the time scale for initiating either plate tectonics or proto-subduction based on plate generation physics. Using the pseudoplastic rheology, \cite{Noack2013} generally find rapid onset times (e.g. $\sim 1$ Myrs) for an Earth-like planet at very low friction coefficients (corresponding to a low yield stress, as is typically needed to generate plate tectonics with pseudoplasticity).  As the friction coefficient increases, the models reach a tipping point where the onset times rapidly increase until the models no longer initiate plate tectonics (e.g. they are in the stagnant lid regime).  This rapid increase in onset time over a narrow range of friction coefficients corresponds to the region of parameter space where these models transition to the stagnant lid regime.  A relatively short onset time when conditions are favorable for plate tectonics, through the use of a small yield stress, is not unexpected because the pseudoplastic rheology, being a simple, instantaneous rheology, has no inherent timescale (material is instantly weakened when the yield stress is reached). However future studies are needed to more fully understand how the onset time scales with Rayleigh number, yield stress, and viscosity (or viscosity ratio), in order to make a more complete comparison between the timescale for initiating plate tectonics with the pseudoplastic rheology versus grain-damage.          

\section{Conclusions}
\label{sec:conclusions}

Scaling laws for convection with grain-damage applied to the Hadean and Eoarchean Earth demonstrate that mantle convection at this time was capable of sustaining proto-subduction and lithospheric mobility, even if the primordial atmosphere was CO$_2$ rich.  However, the plate speed would have been slower, and the lithosphere thicker, because the high mantle temperatures of the Hadean speed up lithospheric grain-growth.  Faster grain-growth leads directly to higher viscosity lithospheric shear zones, which resist plate motion.  The style of subduction in the Hadean was likely different than true plate tectonics; downwellings would be more drip-like than slab-like, and lithospheric deformation would be more diffuse owing to the effects of more rapid lithospheric grain-growth.  However subduction and proto-plate motion would still be maintained.  The conclusion that a sluggish form of proto-subduction was possible in the Hadean is consistent with inferences of subduction from Hadean zircons.  That subduction at the high mantle temperatures (and possibly high surface temperatures) of the Hadean is not truly plate-tectonic may also indicate that the style of subduction inferred from zircons is also distinct from modern day plate tectonics.

To further test the plausibility of Hadean proto-subduction, numerical models of mantle convection immediately after magma ocean solidification are used to constrain the lag time, $t_{lag}$, or the time required for proto-plate tectonics to initiate once convection has begun. Our numerical results show three main phases of evolution as first convection, then proto-plate tectonics, initiate: an initial overturn where the static beginning state first goes unstable; a transient period of stagnant lid convection where damage builds up in the lithosphere; and the initiation of subduction, when sufficient damage has accumulated in lithospheric shear zones. With increasing damage number, $D$, and Rayleigh number, $Ra_0$, the period of stagnant lid convection shrinks as proto-subduction can initiate more rapidly with more efficient grain-damage. Chemical differentiation during magma ocean solidification produces an initially unstable compositional profile.  Including this effect speeds up the initiation of subduction.  The initial overturn happens more rapidly because of the extra buoyancy available to drive flow, and proto-plate tectonics initiates more rapidly because chemically dense material, ``trapped" in the lithosphere after the initial overturn, aids in the development of lithospheric damage.  

A scaling law developed from the numerical models relates the lag time to the damage number, Rayleigh number, and viscosity ratio.  Using this scaling law to estimate the lag time for the early Earth yields a range of  $\approx 100$ Myrs all the way to very short lag times of $\approx 100$ yrs, where proto-subduction would initiate almost concurrently with the beginning of convection. Thus, our numerical and scaling law results indicate that the hypothesis of Hadean subduction, as inferred from Hadean zircon data, is geodynamically plausible.  The initial subduction event involves rapid plate motions and recycling of the lithosphere.  After the first phase of proto-subduction, weak zones inherited from the initial subduction event allow sluggish proto-plate tectonics to persist on the Hadean Earth.  Over time, as both mantle temperature and surface temperature cool, plate speed increases, and modern style plate tectonics develops.  The initial subduction event, and subsequent Hadean proto-subduction, may draw excess CO$_2$ out of the primordial atmosphere, significantly aiding the development of modern style plate tectonics by lowering the surface temperature.


%
%
%
%
%
%

%
%
%
%

\begin{acknowledgments}
This work was supported by NSF award EAR-1135382: Open Earth Systems, and by the facilities and staff of the Yale University Faculty of Arts and Sciences High Performance Computing Center.  We also thank an anonymous reviewer for a thoughtful and thorough review which helped significantly improve the manuscript.
\end{acknowledgments}

%
%
%
%
%
%
%
%
%
%

%




%
%

\end{article}




%
%
%
%
%
%


\end{document}